\documentclass[a4paper,fleqn,twoside,usenatbib]{mnras}

\usepackage{newtxtext,newtxmath}
\usepackage[T1]{fontenc}
\usepackage{ae,aecompl}
\usepackage[toc,page]{appendix}
\usepackage{hyperref}

\usepackage{graphicx}	
\usepackage{amsmath}	

\usepackage{amssymb}	
\usepackage{subfigure} 

\usepackage{cleveref}
\usepackage{booktabs}
\usepackage[table]{xcolor}
\usepackage{array,ragged2e}
\newcolumntype{L}{>{\raggedright\arraybackslash}X}
\usepackage{etoolbox}
\usepackage{tikz,xcolor,hyperref}

\definecolor{lime}{HTML}{A6CE39}
\DeclareRobustCommand{\orcidicon}{
	\begin{tikzpicture}
	\draw[lime, fill=lime] (0,0) 
	circle [radius=0.16] 
	node[white] {{\fontfamily{qag}\selectfont \tiny ID}};
	\draw[white, fill=white] (-0.0625,0.095) 
	circle [radius=0.007];
	\end{tikzpicture}
	\hspace{-2mm}
}

\foreach \x in {A, ..., Z}{\expandafter\xdef\csname orcid\x\endcsname{\noexpand\href{https://orcid.org/\csname orcidauthor\x\endcsname}
		{\noexpand\orcidicon}}
}

\title[Deprojection of barred galaxies]{Deprojection of external barred galaxies from photometry}

\author[B. Tahmasebzadeh et al.]{
	Behzad Tahmasebzadeh,$^{1,2}$\orcidB{} \thanks{behzadtahmaseb@gmail.com,}
	Ling Zhu,$^{1}$\orcidL{} \thanks{Corr author: lzhu@shao.ac.cn;} 
	Juntai Shen,$^{3,4,1}$\orcidJ{} \thanks{jtshen@sjtu.edu.cn.} 
	\newauthor
	Ortwin Gerhard,$^{5}$\orcidO{}
	Yujing Qin$^{6}$\orcidY{}
		\\
		\\
		$^{1}$ Shanghai Astronomical Observatory, Chinese Academy of Sciences, 80 Nandan Road, Shanghai 200030, China \\
		$^{2}$ Department of Astronomy and Space Sciences, University of Chinese Academy
		of Sciences, 19A Yuquan Road, Beijing 100049, China \\
		$^{3}$ Department of Astronomy, School of Physics and Astronomy, Shanghai Jiao Tong University, 800 Dongchuan Road, Shanghai 200240, China \\
		$^{4}$ Key Laboratory for Particle Astrophysics and Cosmology (MOE) / Shanghai Key Laboratory for Particle Physics and Cosmology, Shanghai 200240, China \\
		$^{5}$ Max-Planck-Institut f{\"u}r Extraterrestrische Physik, Gie{\ss}enbachstra{\ss}e 1, 85748 Garching, Germany\\
		$^{6}$ Department of Astronomy/Steward Observatory, 933 North Cherry Avenue, University of Arizona, Tucson, AZ 85721, USA
	}

\pubyear{2021}

\begin{document}
	\label{firstpage}
	\pagerange{\pageref{firstpage}--\pageref{lastpage}}
	\maketitle
   
	
	\begin{abstract}
	The observations of external galaxies are projected to the 2D sky plane. Reconstructing the 3D intrinsic density distribution of a galaxy from the 2D image is challenging, especially for barred galaxies, but is a critical step for constructing galactic dynamical models. Here we present a method for deprojecting barred galaxies and we validate the method by testing against mock images created from an N-body simulation with a peanut-shaped bar. We decompose a galaxy image into a bulge (including a bar) and a disk. By subtracting the disk from the original image a barred bulge remains. We perform multi-Gaussian expansion (MGE) fit to each component, then we deproject them separately by considering the barred bulge is triaxial while the disk is axisymmetric. We restrict the barred bulge to be aligned in the disk plane and has a similar thickness to the disk in the outer regions. 
	The 3D density distribution is thus constructed by combining the barred bulge and the disk. Our model can generally recover the 3D density distribution of disk and inner barred bulge regions, although not a perfect match to the peanut-shaped structure. By using the same initial conditions, we integrate the orbits in our model-inferred potential and the true potential by freezing the N-body simulation. We find that $ 85\% $ of all these orbits have similar morphologies in these two potentials, and our model supports the orbits that generate a boxy/peanut-shaped structure and an elongated bar similar to these in the true potential.
	
\end{abstract}


	
	
	\section{Introduction} \label{intr}
A large fraction of disk galaxies  ($ 30\% $ to $65 \%  $) have bars with various strengths in the central regions \cite[]{Eskridge.2000, Delmestre.2007, Barazza.2008, Aguerri.2009, Gaddoti.2009, Erwin.2018}. In the case of face-on and moderately inclined galaxies, bars appear as non-axisymmetric perturbations in the surface density maps. For the  edge-on or highly inclined galaxies, bars are detectable by particular kinematic signatures, e.g., a positive correlation between mean velocity and the third Gauss-Hermite moment $ h_{3} $ \cite[]{2005.Bureau, 2018.Li}. Bars can redistribute the angular momentum and energy of the disk material, so they drive the morphological evolution of disk galaxies \cite[]{1985.Weinberg, 1998.Debattista, 2003.Athanassoula, KK.2004}. They are well known to play a major role in the secular galaxy evolution \cite[]{1993.Friedli, 2005.Sheth, 2011.Masters, 2011.Gadotti}. \par A bar has been included in the dynamical models for Milky Way by the Schwarzschild orbit-superposition method \cite[]{1996.Zhao, Mao.2012} and the made-to-measure (M2M) method \cite[]{2013.Long, 2013.Hunt, 2015.Portail, Portail.2017}. There are several commonly used implementations of the Schwarzschild's orbit-superposition method for external galaxies by taking different assumptions of their geometries: they could be spherical \cite[]{Richstone.1985, 2013.Breddels, 2017.Kowalczyk}, axisymmetric \cite[]{1999.Cretton, Gebhardt.2000, 2004.Valluri} or triaxial \cite[]{Cappellari.2006, Bosch.2008, ling.1018}. However, bar shapes and the figure rotation are not included explicitly in these models.
\par A major goal of dynamical modeling is to obtain the underlying mass distribution, including the central supermassive black hole mass, the stellar mass profile, and the dark matter halo profile. However, without including the bar properly, these results could be significantly biased  \cite[]{Brown.2013}. Dynamical modeling of external barred galaxies are still in the early stage.  \cite{Bla.2018} made a triaxial bulge/bar/disk M2M model for M31 taking an N-body model which generally matches the bulge properties of M31 as an initial condition of the M2M algorithm. A bar has been included in the recently developed Schwarzschild code, SMILE \cite[]{Vasiliev.2013} and FORSTAND \cite[]{Vasiliev.2019c}, which are, however, only applied to mock data created from a simulation by using its real 3D density distribution. Estimating the 3D density distribution of a real barred galaxy is a key step still missing before we can create proper dynamical models to a real barred galaxy from observations in a general sense. 
\par
It is non-trivial to obtain the intrinsic 3D density distributions from the 2D images on the observational plane for triaxial systems, and even for axisymmetric systems when not observed edge-on \cite[]{Rybicki}. Several approaches are developed to make the 3D deprojection feasible. One of such approaches is to consider ellipsoid luminosity density profiles for stellar systems with ellipsoidal radius of  $ m = \sqrt{x^{2} + \frac{y^{2}}{p^{2}} + \frac{z^{2}}{q^{2}}} $ which requires the axial ratios  $ q \leq p \leq 1 $. The parameters could then be derived by fitting the projected model to the observed image  \cite[]{Contopoulos.1956, Stark.1977, binny.1985}.  However, these methods are not able to reproduce isophotal twists and ellipticity variations along the radius. Non-parametric deprojections are also attempted \cite[]{Magorian.1999, Gerhard.2002}, but they are complicated and require considerable time to converge. Multi-Gaussian expansion (MGE) \cite[]{Bendinelli.1991, Emsellem.1992, Emsellema.1994, cappellari.2002} is an efficient method to describe the surface brightness of a 2D image and deproject it to the 3D luminosity density distribution. It can reproduce isophotal twists and ellipticity variations along the radius for a triaxial rigid body. In addition, the gravitational potential and force components can be computed based on the 3D density distribution derived from this method \cite[]{Bosch.2008}. MGE has been widely used in dynamical models for different types of galaxies \cite[e.g.,][]{Emsellemb.1994, Cretton.1999, Cappellari.2008, Zhu.2018}, but not yet including bars explicitly.
\par 
A popular method of deprojecting barred galaxies is to describe their face-on view by analytical, image stretching, or Fourier-based methods \cite[]{Gadotti.2007, Noordermeer.2007, LiYu.2011}, then adopting a scale height for the vertical profile to construct the 3D density distribution (e.g., \cite{Weiner.2001}).  This approach works for galaxies with moderate inclination angles ( $< \hspace{0.02cm} 60^{\circ}$) and the vertical thickness of the bar is a significant source of uncertainties \cite[]{Zou.2014, Fragkoudi.2015}. The 3D intrinsic shape of bars can also be obtained with some statistical approaches. \cite{Abreu.2010, Abreu.2018} developed an approach to derive the intrinsic shape of bulges/bar based on the geometric information extracted from  2D photometric decomposition. They assume a bulge/bar is a triaxial ellipsoid that shares the same equatorial plane as an oblate disk. They use this method to estimate the intrinsic axes ratios of $ 83 $ barred galaxies from CALIFA survey, and they find that $ 68 \% $ of bars in their sample are prolate-triaxial ellipsoids and $ 32 \% $ are oblate-triaxial ellipsoids.
\par
In this paper, we present an efficient way of deprojecting barred galaxies based on the MGE algorithm, and validate the method by applying it to mock images created from a simulation.  We first create mock images with different projection angles as described in \S \ref{s-moc}. We introduce our deprojecting approach for barred galaxies in \S \ref{m-de}. For verification of our model, in \S \ref{m-pot} we calculate the potential, force and analyze the orbital structures according to the 3D density distribution from our model, and compare with those from the original simulation. In \S \ref{s-con}, we summarize and list the main conclusions of our work.

\section{the mock data} \label{s-moc}
We use an influential N-body simulation of a Milky Way-like galaxy from \cite{shen.2010}.  It has a central bar-shaped structure that matches many observed properties of the Milky Way \cite[]{shen.2015}. The total stellar mass of the simulation is $ 4.25 \times 10^{10} M_{\odot} $ with $ 10^{6} $ equal-mass particles. A rigid pseudo-isothermal DM halo potential is adopted  $\Phi=\frac{1}{2} V_{\mathrm{0}}^{2} \ln \left(1+\frac{r^{2}}{R_{\mathrm{c}}^{2}}\right)$, in which the scale velocity and scale radius are $V_{\mathrm{0}}=250  \hspace{.05cm} \mathrm{km}\mathrm{s}^{-1}$ and $R_{c}=15 \hspace{.05cm} \mathrm{kpc}$,  respectively. The bar has a half-length of $  \sim  \hspace{.03cm} 4 \hspace{.05cm} \mathrm{kpc} $ and rotates with a pattern speed of $ \Omega_{ \mathrm{p}} \sim 39 \hspace{.08cm} \mathrm{km \hspace{.04cm} s^{-1} \hspace{.04cm} kpc^{-1} }$ (corotation radius $  \sim  \hspace{.03cm} 4.7 \hspace{.05cm} \mathrm{kpc} $).
The end-to-end separation between the outer two edges of the $ \mathrm{X} $-shaped structures is	$ \sim \hspace{.03cm} 4 \hspace{.05cm} \mathrm{kpc} $ along the major axis and $\sim  \hspace{.03cm} 2.4 \hspace{.05cm} \mathrm{kpc} $ along the vertical minor axis \cite[]{li.2012}.

\par 
 Throughout the paper, we use the coordinate $ (x, y, z) $ to describe the intrinsic 3D structure, where $ x $, $ y $, $ z $ are aligned with the long, intermediate and short axes of the galaxy. While we use the coordinate $ (x^{\prime}, y^{\prime}) $ to describe the projected structure to the 2D observational plane, and $ z^{\prime} $ is along the line-of-sight. 
The orientation of a projection is specified by the viewing angles $ (\theta,\varphi,\psi) $. $ \theta $ and $ \varphi $ give the orientation of the line-of-sight with respect to the principal axes of the object. For instance, projections along the intrinsic major, intermediate, and minor axes correspond respectively to $(\theta=90^{\circ},\varphi=0^{\circ}) $, $(\theta=90^{\circ},\varphi=90^{\circ}) $ and $(\theta=0^{\circ},\varphi=0^{\circ}, ... , 90^{\circ}) $. The angle $ \psi $ is required to determine the rotation of the object around the line of sight (see Fig. 2 in \cite{Zeeuw.1989}).	
\par The two coordinate systems are related as \cite[]{binny.1985}:	
\begin{equation}
\left(\begin{array}{l}{x^{\prime}} \\ {y^{\prime}} \\ {z^{\prime}}\end{array}\right)=\mathbf{R} \cdot \mathbf{P} \cdot\left(\begin{array}{l}{x} \\ {y} \\ {z}\end{array}\right),
\end{equation}	
where matrix $ \mathrm{P} $ is responsible for the projection onto the sky plane defined as:
\begin{equation}
{\mathbf{P}=\left(\begin{array}{ccc}{-\sin \varphi} & {\cos \varphi} & {0} \\ {-\cos \theta \cos \varphi} & {-\cos \theta \sin \varphi} & {\sin \theta} \\ {\sin \theta \cos \varphi} & { \sin \theta \sin \varphi } & {\cos \theta}\end{array}\right)},
\end{equation}
and matrix $ \mathrm{R} $ expresses the rotation around the line-of-sight by angle $ \psi $:
\begin{equation}
{\mathbf{R}=\left(\begin{array}{ccc}{\sin \psi} & {-\cos \psi} & {0} \\ {\cos \psi} & {\sin \psi} & {0} \\ {0} & {0} & {1}\end{array}\right)}.
\end{equation} 
\par
To simplify the description of multiple components of a galaxy, we always align the major axis of the disk with $ x^{\prime} $ axis and minor with $ y^{\prime} $. Note that this could be different from the natural coordinate, i.e. $ y^{\prime} $ aligning with north, when dealing with real observational data. Here $\psi$ is defined from the $ y^{\prime} $ axis, thus we always have $\psi_{\mathrm{ disk}} = 90^{\circ} $ for the disk aligned in this way.
\par To create a mock image, we project the simulation snapshot to the observational plane with certain viewing angles. We put it at a distance of $ 41 \hspace{.05cm} \mathrm{Mpc} $ and then create a surface brightness map with a spatial resolution of  $ 1 \hspace{.05cm} \mathrm{arcsec} \hspace{.08cm} \mathrm{pixel}^{-1} $. At this distance, $ 1 \hspace{.05cm} \mathrm{kpc}$ equals $ 5 \hspace{.05cm} \mathrm{arcsec}$. We produce several mock images from the simulation adopting different viewing angles of $ \theta $ and $ \varphi $ as listed in Table \ref{table:mockt}, which are labeled as true values. To keep the disk major axis aligned with $ x^{\prime} $-axis,  all of our mock galaxies have $\psi^{\mathrm{true}}=90^{\circ} $. Each mock image will be taken as an independent galaxy from observation. In the following sections, we will illustrate our deprojection model with the mock galaxy $I_{1}$.  
\begin{table}
	 \centering
     \small
	\begin{tabular}{p{0.30\linewidth}p{0.30\linewidth}p{0.15\linewidth}}
	\hline
	\toprule
	Name    & $\theta^{\mathrm{true}} $  $(^{\circ}) $ & $  \varphi^{\mathrm{true}} $ ($ ^{\circ} $)  \\
	\midrule  
	$ I_{1} $      & 60    & -45     \\ 
	$ I_{2} $      & 60    & -90     \\ 
	$ I_{3} $      & 60    & 0      \\ 
	$ I_{4} $      & 80    & -45     \\
	$ I_{5} $      & 80    & -90     \\ 
	$ I_{6} $      & 80    & 0      \\ 
	\bottomrule
   \end{tabular}
\parbox{\columnwidth}{\caption{Mock data sets with different viewing angles of $ \theta^{\mathrm{true}} $ and $ \varphi^{\mathrm{true}}$. Inclination angle $ \theta^{\mathrm{true}} $ is the orientation of the disk,  $ \theta^{\mathrm{true}}=90^{\circ} $ means edge-on and $ \theta^{\mathrm{true}}=0^{\circ} $ means face-on. $ \varphi^{\mathrm{true}}$ describes the orientation of the bar, $ \varphi^{\mathrm{true}} = 90^{\circ}$ means side-on and $ \varphi^{\mathrm{true}} = 0^{\circ}$ means end-on. To keep the disk major axis aligned with $ x^{\prime} $-axis, we have $\psi^{\mathrm{true}}=90^{\circ} $ for all cases.}
\label{table:mockt}}
\end{table}

\section{Derojection} \label{m-de}
Here, we give a step-by-step description of our method. We first decompose the galaxy to a bulge and a disk by using GALFIT \cite[]{Peng.2010}. Secondly, we fit the bulge and the disk using 2D MGE separately. Then, we deproject each component individually from 2D MGE to 3D MGE. Finally, the deprojected galaxy is simply the sum of the axisymmetric disk and the triaxial bulge. Meanwhile, the bulge major axis is restricted to be aligned in the disk plane.
\par
This method allows different intrinsic shapes for the two components. And we have the freedom to align the bulge in the disk plane with different position angles.

\subsection{2D bulge-to-disk decomposition} \label{s-decom}
We use GALFIT 3.0.5 \footnote{\url{https://users.obs.carnegiescience.edu/peng/work/galfit/galfit.html}}  to decompose the surface brightness map of our mock galaxy into 2D elliptical bulge and disk. Poisson noise is taken as the uncertainty of the image, which is used to weight the data points in GALFIT fitting.
\par We use a standard exponential profile to describe the surface brightness of a disk:
\begin{equation}
\label{disksb}
\Sigma(r^{\prime})=\Sigma_{0} \exp \left(-\frac{r^{\prime}}{r_{s}}\right),
\end{equation}  
where $ r_{s} $ is the scale length of disk and  $ r^{\prime}=\left(x^{\prime 2}+\frac{y^{\prime 2}}{q_{d}^{2}} \right)^{\frac{1}{2}} $, $ (x^{\prime}, y^{\prime}) $ are aligned with the major and minor axis of the disk, and $ q_{d} $ is the axial ratio.

\par We use a S\'{e}rsic function to describe the surface brightness of an elliptical bulge:
	\begin{equation}
	\label{sersic}  
	\Sigma(r^{\prime})=\Sigma_{e} \exp \left[-\kappa\left(\left(\frac{r^{\prime}}{r_{e}}\right)^{\frac{1}{n}}-1\right)\right],
	\end{equation}
 Where $ r_{e} $ is the effective radius (or half-light radius), $ \Sigma_{e} $ is the surface brightness at the effective radius $ r_{e} $, and $ n $ is called the "S\'{e}rsic index", which controls the concentration of the profile. When $ n $ is small, it has a shallow inner profile and a steep truncation at a large radius. Inversely, When $ n $ is large, it has a steep inner profile and a highly extended outer wing.  $ \kappa $ is not a free parameter, it coupled to $ n $ by $\gamma\left(2 n ; \kappa\right)=\frac{1}{2} \Gamma(2 n)$, $\Gamma$ and $ \gamma $ are  the Gamma function and lower incomplete Gamma function, respectively. Here $ r^{\prime} $ is defined as:
 \begin{equation}
 \label{2dsb} 
 r^{\prime}=\left(x_{p}^{2}+\frac{y_{p}^{2}}{q_{b}^{2}} \right)^{\frac{1}{2}}.
 \end{equation}  
  $ q_{b} = b/a $, where $ b $ and $ a $ are the minor and major axis of the elliptical bulge in the observational plane. $ x_{p} $ and $ y_{p} $ are aligned with the major and minor axis of the elliptical bulge:
   \begin{equation}
  \begin{split}
  & x_{p} =  \hspace{.16cm}x^{\prime} \cos (90- \mathrm{\psi}^\mathrm{proj}_\mathrm{bar}) + y^{\prime} \sin (90- \mathrm{\psi}^\mathrm{proj}_\mathrm{bar})  
  \\ & y_{p} = - x^{\prime} \sin (90- \mathrm{\psi}^\mathrm{proj}_\mathrm{bar}) + y^{\prime} \cos (90- \mathrm{\psi}^\mathrm{proj}_\mathrm{bar})  
  \end{split}
  \end{equation}
where $ \mathrm{\psi}^\mathrm{proj}_\mathrm{bar} $ is the position angle of the elliptical bulge measured counter-clockwise from the  $y^{\prime}$-axis.
\par
\begin{figure}%
	\includegraphics[width=\columnwidth]{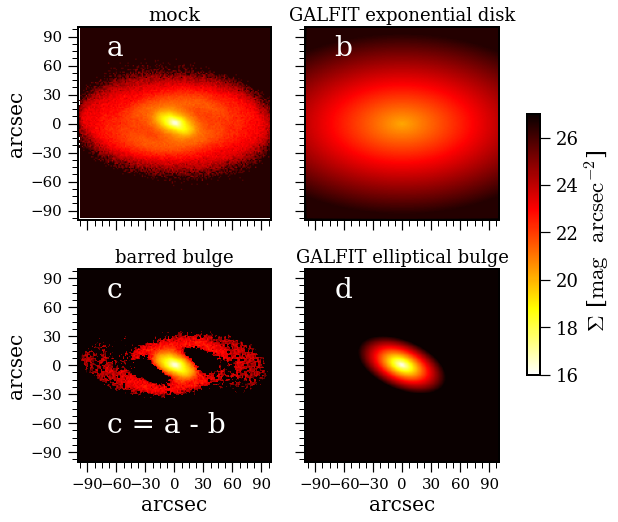}
	\caption{2D surface brightness profiles of:
		 (a) original mock galaxy $ I_{1}$, (b) GALFIT exponential disk of the best-fit model, (c) barred bulge which is obtained by subtracting the GALFIT disk from the original image, (d) the GALFIT elliptical bulge (S\'{e}rsic profile).}%
	\label{gal-dec}%
\end{figure}

\begin{figure}
	\includegraphics[width=8cm]{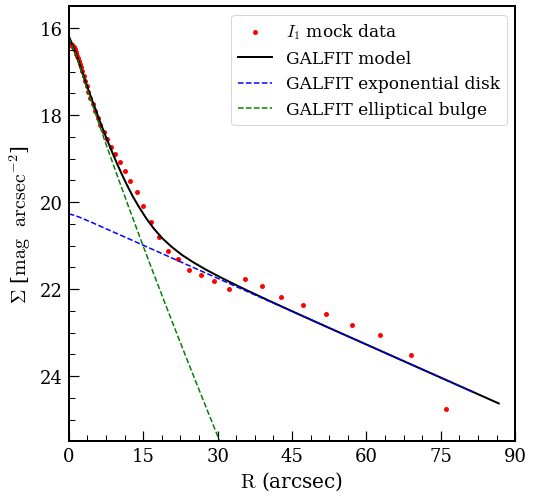}
	\caption{ Surface brightness profile along the major axis. The blue and green dashed curves are the exponential disk and the elliptical bulge from GALFIT, while the black solid curve indicates the sum of them.  The red dots are the data of the mock galaxy $ I_{1}$.}
	\label{1Dsur}
\end{figure}

 Fig. \ref{gal-dec} presents the 2D surface brightness profile of (a) the mock galaxy $ I_{1} $, (b) GALFIT exponential disk, (c) barred bulge, which is the residual of subtracting GALFIT disk from the original image, (d) GALFIT elliptical bulge. The radial profiles of the surface brightness along the major axis is presented in Fig. \ref{1Dsur}. The sum of an exponential disk and an elliptical bulge gives a reasonable fit to the image. although the model is slightly dimmer in the regions of $40 \lesssim R \lesssim 70$  $\mathrm{arcsec}$, and brighter at $ R \gtrsim 70$  $\mathrm{arcsec}$.  Our best-fit parameters are $r_{s}=20.90$ $\mathrm{arcsec}$, $ \Sigma_{0}=11.63 $ $ \mathrm{mag}$  $\mathrm{arcsec}^{-2} $, $q_{d}=0.51$ for the disk component and $n=1.093$, $r_{e}=5.88$ $\mathrm{arcsec}$, $ \Sigma_{e}=11.69 $ $ \mathrm{mag} $  $\mathrm{arcsec}^{-2} $, $q_{b}=0.46$, and $\mathrm{\psi}^\mathrm{proj}_\mathrm{bar}=71^{\circ}$ for the elliptical bulge. Note that we have $ \psi_\mathrm{disk} = 90^\circ $, thus the position angle of the bar is $ -19^{\circ} $ different from the disk. 
  
\subsection{2D MGE fit to the bulge and disk} \label{mgefit}
 We then separately fit 2D MGEs to (i) the 2D surface decomposed disk and (ii) the residual barred bulge mock image obtained by subtracting the disk model from the original mock image. This is done with the $\mathrm{MgeFit}$ software from \cite{cappellari.2002}  \footnote{\url{http://www-astro.physics.ox.ac.uk/~mxc/software/}}. The MGE describes the surface brightness (in the unit of $ \mathrm{L_{sun}} \mathrm{pc}^{-2} $ converted from surface brightness in the unit of $ \mathrm{mag}$  $\mathrm{arcsec}^{-2} $) written as \cite[]{cappellari.2002}:
\begin{equation}\label{MGESB}
\mathbf{\Sigma}\left(R^{\prime}, \theta^{\prime}\right)=\sum\limits_{\substack{i=0 }}^N \frac{L_{j}}{2\pi \sigma'^{2}_{j}q'_{j}}\exp \left[-\frac{1}{2\sigma'^{2}_{j}} \left( x_{j}^{\prime 2}+\frac{y_{j}^{\prime 2}}{q'^{2}_{j}} \right) \right],
\end{equation}
with ($ j $ refers to each Gaussian). 
\begin{equation}
{x_{j}^{\prime}=R^{\prime} \sin \left(\theta^{\prime}-\psi_{j}^{\prime} \right)}, \hspace{0.5cm} {y_{j}^{\prime}=R^{\prime} \cos \left(\theta^{\prime}-\psi_{j}^{\prime} \right)},
\end{equation}
where $ (R^{\prime}, \theta^{\prime}) $ are polar coordinates in the sky plane. $ L_{j} $ indicates the observed total luminosity, $ q_{j}^{\prime}$ is the projected flattening  and we assume $0 \leqslant q_{j}^{\prime} \leqslant 1$,  $\sigma_{j}^{\prime}$ is the scale length along the projected major axis, and $\psi_{j}^{\prime}$ is the position angle measured counter-clockwise from the $y^{\prime}$-axis to the major axis of each Gaussian component. We denote:
\begin{equation}
\label{pstw}
\psi_{j}^{\prime}=\psi+\Delta \psi_{j}^{\prime}
\end{equation}
where $ \Delta \psi_{j}^{\prime} $ is the isophotal twist of each Gaussian that can be measured directly.
\par
\begin{figure}%
	\subfigure[][]{%
		\label{fig:mged}%
		\includegraphics[width=4.0cm]{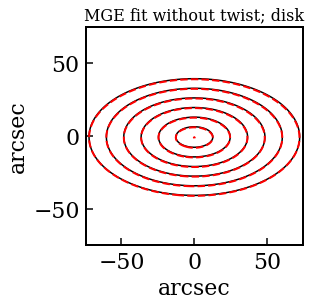}}
	\hspace{0.17cm}
		\subfigure[][]{%
		\label{fig:mgeb}%
		\includegraphics[width=4.3cm]{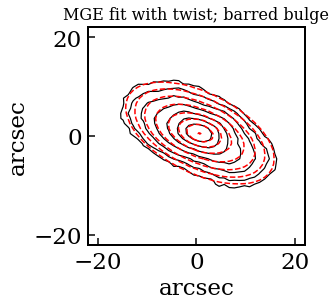}}
	\caption[A set of four subfigures.]{
		(\subref{fig:mged}) The contours of the disk image (black solid line), over-plotted with contours of the best-fit MGE without twist between different Gaussians (red dashed line). 
		(\subref{fig:mgeb}) The MGE fit including twist for the barred bulge.}%
	\label{gal-mge}%
\end{figure}

 
For the disk component, we always have $ \psi_{j}^{\prime}=\psi_{\mathrm{disk}}= 90 ^{\circ}$ by aligning its major axis with $ x^{\prime} $, and $ \Delta \psi_{j}^{\prime} = 0 ^{\circ}$ for all Gaussian components by assuming an axisymmetric oblate shape. While for the barred bulge component which may be triaxial, the isophotal twist of each Gaussian component is allowed to be different. Here we measure twists with respect to the disk position angle ($ \Delta \psi_{j}^{\prime}  = \psi_{j}^{\prime}-\psi_{\mathrm{disk}} $). Details of fitting are presented in Table  \ref{table:diskmgep} and \ref{table:barmgep} respectively. Results are shown in Fig. \ref{gal-mge}. There are weak signals of possible spurs which are offset from the major axis of the inner region of the barred bulge in Fig. \ref{fig:mgeb} (black solid line contours). The offset-spurs are isophotal signatures corresponding to the vertically boxy/peanut part of the bar in moderately inclined barred galaxies \cite[]{Erwin.2016}.

\begin{table}
	\centering
	\footnotesize
	\begin{tabular}{p{0.15\linewidth}p{0.15\linewidth}p{0.15\linewidth}p{0.15\linewidth}p{0.15\linewidth}}
		\hline
		\toprule
		$ j $    &  $ L_{j}$ $(\mathrm{L_{\odot}pc^{-2}}) $ &  $ \sigma^{\prime}_{j} $ $ (\mathrm{arcsec}) $    & $ q^{\prime}_{j} $   &  $ \Delta \psi_{j}^{\prime} $ $(^{\circ})$   \\
		\midrule
		$ 1 $      & 39.380    & 2.400    &   0.547   &   0  \\
		$ 2 $      & 42.381    & 5.671    &   0.558   &   0  \\
		$ 3 $      & 48.220    & 9.592    &   0.554   &   0  \\
		$ 4 $      & 54.451    & 14.300   &   0.555   &   0  \\
		$ 5 $      & 74.615    & 21.589   &   0.557   &   0  \\	
		$ 6 $      & 55.593    & 34.948   &   0.551   &   0  \\	
		$ 7 $      & 10.840    & 54.956   &   0.568   &   0  \\				
		\bottomrule
	\end{tabular}
	\\
	\parbox{\columnwidth}{\caption{Details of MGE fit for the disk component in Fig. \ref{fig:mged}. $ j $ is the number of each individual Gaussian for which,  $ L_{j}$ is the central flux in the unit of $(\mathrm{L_{\odot}pc^{-2}})$, $ \sigma^{\prime}_{j} $ presents the size in the unit of $ (\mathrm{arcsec}) $, $q^{\prime}_{j}$ indicates the flattening. Position angles of all disk Gaussians components are fixed to be $ 90^{\circ} $. }
		\label{table:diskmgep}}
\end{table}

\begin{table}
	 \centering  
	 \footnotesize		
	    \begin{tabular}{{p{0.15\linewidth}p{0.15\linewidth}p{0.15\linewidth}p{0.15\linewidth}p{0.15\linewidth}}} 
		\hline
		\toprule
		$ j $    &  $ L_{j}$ $(\mathrm{L_{\odot}pc^{-2}}) $ &  $ \sigma^{\prime}_{j} $ $ (\mathrm{arcsec}) $    & $ q^{\prime}_{j} $   &  $ \Delta \psi_{j}^{\prime} $  $(^{\circ})$  \\
		\midrule   
		$ 1 $      & 1666.431    & 0.864    &   0.544  &    -27.796  \\
		$ 2 $      & 7924.190    & 2.136    &   0.593  &    -17.000  \\
		$ 3 $      & 1256.558    & 4.848    &   0.480  &    -17.000  \\
		$ 4 $      & 1907.776    & 6.499    &   0.480  &    -28.000  \\
		$ 5 $      & 117.221     & 10.212   &   0.600  &    -28.000  \\		
		\bottomrule
     	\end{tabular}  
     \\
	\parbox{\columnwidth}{\caption{Details of MGE fit for the barred bulge component in Fig. \ref{fig:mgeb}. $ \Delta \psi_{j}^{\prime} $ shows the isophotal twist with respect to the disk. }
	\label{table:barmgep}}
\end{table}

\subsection{Construction of 3D MGE density} \label{dep-se}
  We first deproject each component from 2D MGE surface brightness to 3D intrinsic density distribution separately, then add the 3D density distributions of the two components together for a whole galaxy. An oblate shape and a triaxial ellipsoid are used to describe disk and barred bulge, respectively. The intrinsic coordinate system denoted by $ (x, y, z) $ is aligned with the galaxy's principal axes. 
\par The triaxial MGE luminosity density is defined as \cite[]{cappellari.2002}:

\begin{equation}\label{MGESB2}
\rho(x,y,z)=\sum\limits_{\substack{j=0 }}^N \frac{L_{j}}{(\sigma_{j} \sqrt{2\pi })^{3} q_{j} p_{j}}\exp \left[-\frac{1}{2\sigma^{2}_{j}} \left( x^{2}+\frac{y^{2}}{p^{2}_{j}}+\frac{z^{2}}{q_{j}^{2}} \right) \right], 
\end{equation}
where $ N $ is the number of Gaussian functions. $ p_{j} = B_{j}/A_{j} $ and $ q_{j} = C_{j}/A_{j} $ are the axial ratios. $ A_{j} $, $ B_{j} $, and $ C_{j} $ are the intrinsic major, intermediate, and minor axes of the Gaussians.

\subsubsection{Deprojection of barred bulge: triaxial case} \label{teri-d}
A deprojection could be understood as a transformation between two coordinates as following:
\begin{equation}
\left(\begin{array}{l}{x} \\ {y} \\ {z}\end{array}\right)=\mathbf{R^{-1}} \cdot \mathbf{P^{-1}} \cdot\left(\begin{array}{l}{x^{\prime}} \\ {y^{\prime}} \\ {z^{\prime}}\end{array}\right).
\end{equation}

\par From parametric Eq. (\ref{MGESB}), we have the parameters $ (L_{j}, q'_{j} ,\sigma '_{j}, \Delta \psi_{j}^{\prime} ) $ for each Gaussian from observations. Given a set of viewing angles $ (\theta, \varphi, \psi) $, we can infer the intrinsic quantities $ (\sigma_{j}, p_{j}, q_{j}) $ through the following equations \cite[]{Bosch.2008}:
\begin{equation}\label{int-shape1}
1-q_{j}^{2}=\frac{\delta_{j}^{\prime}\left[2 \cos 2 \psi_{j}^{\prime}+\sin 2 \psi_{j}^{\prime}(\sec \theta \cot \varphi-\cos \theta \tan \varphi)\right]}{2 \sin ^{2} \theta\left[\delta_{j}^{\prime} \cos \psi_{j}^{\prime}\left(\cos \psi_{j}^{\prime}+\cot \varphi \sec \theta \sin \psi_{j}^{\prime}\right)-1\right]},
\end{equation}
\begin{equation}
\label{int-shape2}
p_{j}^{2}-q_{j}^{2}=\frac{\delta_{j}^{\prime}\left[2 \cos 2 \psi_{j}^{\prime}+\sin 2 \psi_{j}^{\prime}(\cos \theta \cot \varphi-\sec \theta \tan \varphi)\right]}{2 \sin ^{2} \theta\left[\delta_{j}^{\prime} \cos \psi_{j}^{\prime}\left(\cos \psi_{j}^{\prime}+\cot \varphi \sec \theta \sin \psi_{j}^{\prime}\right)-1\right]},
\end{equation}
\begin{equation}
\label{int-shape3}
u_{j}=\frac{1}{q_{j}^{\prime}} \sqrt{p_{j}^{2} \cos ^{2} \theta+q_{j}^{2} \sin ^{2} \theta\left(p_{j}^{2} \cos ^{2} \varphi+\sin ^{2} \varphi\right)}.
\end{equation}
Where $u_{j}=\sigma_{j}^{\prime} / \sigma_{j}$, $ \delta'^{2}_{j}=1-q'^{2}_{j} $, and $\psi_{j}^{\prime}=\psi+\Delta \psi_{j}^{\prime}$. Note that, in principle, the absolute value of the position angle $ \psi $ of a triaxial structure could be a free parameter, however, the relative difference of position angle of the Gaussians $ \Delta \psi_{j}^{\prime} $ are measured directly from the MGE fitting.
\par All Gaussians are supposed to have the same viewing angles $ (\theta, \varphi, \psi) $, the allowed viewing angles is thus the intersection of these allowed for each Gaussian component. The allowed range of viewing angles are restricted by the Gaussian with the minimum of flattening $q^{\prime}_{j}$ and the maximum difference of twist $\Delta \psi_{j}^{\prime}$ among the Gaussians. \cite[]{cappellari.2002, Bosch.2008}. To avoid non-physical restriction on the allowed viewing angles caused by a particular Gaussian, we avoid the Gaussians with too small  $q^{\prime}_{j}$ and too large $\Delta \psi_{j}^{\prime}$ difference, when the error does not change significantly during the MGE fitting (see Fig.1 in \cite{Bosch.2008}).

\par We emphasize that if we consider the barred bulge and disk as one rigid body, deprojection is impossible due to the large difference between $\Delta \psi_{j}^{\prime}$ of the barred bulge Gaussians and disk Gaussians. Therefore we need to deproject the barred bulge and disk separately.

\subsubsection{Deprojection of disk: axisymmetric case} \label{axi-d}
We consider the axisymmetric oblate shape for the disk. We always align the disk major axis with $ x^{\prime} $-axis, thus we have $\psi_{\mathrm{disk}} = 90^{\circ}$ and $ \Delta \psi_{j}^{\prime} = 0 $ for all Gaussians of the disk.  For an axisymmetric oblate system $\varphi $ is irrelative, therefore Eqs. \ref{int-shape1} and \ref{int-shape2} are simplified to:
\begin{equation}\label{int-disk}
q^{2}_{j}=\frac{q'^{2}_{j}-\cos^{2}\theta}{\sin^{2}\theta}, \hspace{.3cm}  \hspace{.3cm} p_{j} =1.
\end{equation}
and 
\begin{equation}\label{siq-disk}
\sigma_{j} = \sigma^{\prime}_{j},
\end{equation}
where $ (\theta > 0^{\circ}) $ is the inclination angle of the disk, $ q_{j} $ is the intrinsic flattening and $ q'_{j} $ is the observed flattening of each Gaussian component in the disk. Axisymmetric MGE deprojection above is only valid up until $ (\cos^{2}\theta < q'^{2}_{j}) $ for all Gaussian components. It means the minimum inclination is imposed by the flattest Gaussian in an axisymmetric MGE fit. 
\par In addition, the intrinsic flattening $ q $ of disks in late-type galaxies have a narrow distribution centered at $ q\sim 0.26 $ \cite[]{Rodr.2013}. We can roughly derive  the inclination of the disk with its observed flattening $ q_\mathrm{obs} $, by assuming an intrinsical flattening of $ 0.26 $ from Eq.\ref{int-disk}. We denote the disk inclination angle derived in this way as $ \theta_\mathrm{disk}^\mathrm{derive} $.

\subsection{Allowed viewing angles} \label{degen}
\begin{figure*}%
	\centering
	\subfigure{%
		\includegraphics[width=18cm]{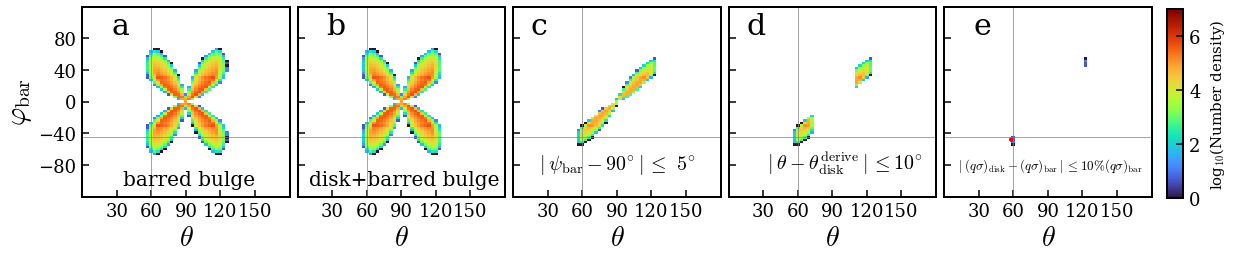}}%
	\caption[]{
		The parameters space of allowed viewing angles  $\theta$ vs. $\varphi$ for the deprojection of barred bulge + disk. Plots from left- to right-hand side: $\mathrm{a)}$ all the allowed viewing angles of the barred bulge, $\mathrm{b)}$ orientations which are also allowed for the disk, $ \mathrm{c)} $ the additional constraint of $| \psi_{\mathrm{bar}} - 90^{\circ}  |  \leq 5^{\circ} $, $ \mathrm{d)} $ the additional constraint of $| \theta - \theta_{\mathrm{disk}}^{\mathrm{derive}} |  \leq 10^{\circ} $, $ \mathrm{e)} $ add the last constraint that outer barred bulge and the inner disk share the similar scale height. The gray lines show the true value of $ \theta^{\mathrm{true}}=60^{\circ} $ and $ \varphi^{\mathrm{true}}=-45^{\circ} $ for the mock galaxy $ I_{1} $. Red dot in last panel indicates a model with ($ \theta=58^{\circ}$, \hspace{0.05cm} $\varphi_{\mathrm{bar}}=-47^{\circ}$, \hspace{0.05cm} $ \psi_{\mathrm{bar}}=92^{\circ} $) from the final allowed angles. }%
	\label{constraint}%
\end{figure*}

  Fig. \ref{constraint} shows the allowed viewing angles ($ \theta $ vs.  $ \varphi $) of the galaxy by combining the disk and barred bulge. In Fig. \ref{constraint}a, we show the allowed regions of viewing angles for the barred bulge, which are intersections of the allowed viewing angles of each individual Gaussian in the barred bulge.
 \par Then, we consider the major axis of the barred bulge is aligned in the disk plane. Thus, the inclination angle of the barred bulge should be the same as the disk. This combination narrows down the allowed inclination angle $ \theta $ as shown in Fig. \ref{constraint}b.
 \par The intrinsic position angle $ \psi $ of an isolated triaxial system is unknown as discussed in \cite{Bosch.2008}. However, in our case, we have a reference disk which is aligned as $\psi_{\mathrm{disk}} = 90^{\circ}$. For the barred bulge fixed in the disk plane, we restricted it to be:
 \begin{equation}\label{const2}
 | \psi_{\mathrm{bar}}  - 90^{\circ}  |  \leq 5^{\circ}.
 \end{equation}
Note that it is $ \psi_{j}^{\prime}=\psi_{\mathrm{bar}}+\Delta \psi_{j}^{\prime} $ that goes into Eqs. (\ref{int-shape1}-\ref{int-shape3}), the large isophotal twist between the barred bulge and disk is considered in $ \Delta \psi_{j}^{\prime} $ (see Table \ref{table:barmgep}).

\par We can further constrain the inclination angle $ \theta $ with the disk as shown in Fig. \ref{constraint}d, based on what we roughly derived in Section \ref{axi-d}:
\begin{equation}\label{const1}
| \theta - \theta_{\mathrm{disk}}^{\mathrm{derive}} |  \leq 10^{\circ}, 
\end{equation}
\par The last constraint we consider refers to the scale height of the disk and barred bulge. We consider similar thickness of the disk and the barred bulge at outer regions of the barred bulge:
\begin{equation}\label{const3}
| (\sigma q)_{\mathrm{disk}}-(\sigma q)_{\mathrm{bar}} |  \leq 10 \hspace{0.02cm} \% \hspace{0.1cm}  (\sigma q)_{\mathrm{bar}}.
\end{equation}
  Where  $ (\sigma q)_{\mathrm{bar}} $,  and $ (\sigma q)_{\mathrm{disk}} $ indicate size and flattening of the outermost barred bulge Gaussian and a Gaussian of the disk that has a roughly similar size. We consider $ (\sigma q)_{\mathrm{bar}} $ of outer most Gaussian of the barred bulge (5th Gaussian in Table. \ref{table:barmgep}) to be close to $ (\sigma q)_{\mathrm{disk}} $ of the 4th Gaussian of the disk (in Table. \ref{table:diskmgep}).  The allowed regions of viewing angles after imposing all the constraints are shown in Fig. \ref{constraint}e.

\subsection{3D density distribution} \label{3dmap}

\begin{figure*}\label{densm}
	\subfigure[][]{%
		\label{fig:dent1}%
		\includegraphics[width=5.8cm]{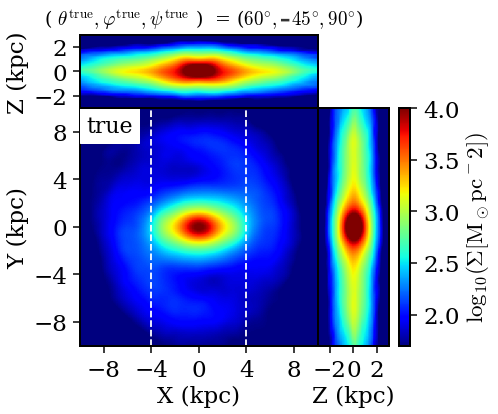}}
	\subfigure[][]{%
		\label{fig:dend}%
		\includegraphics[width=5.7cm]{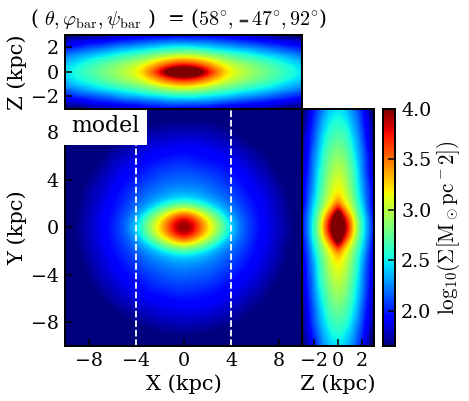}}
		\hspace{8pt}%
	\subfigure[][]{%
		\label{fig:res}%
		\includegraphics[width=5.6cm]{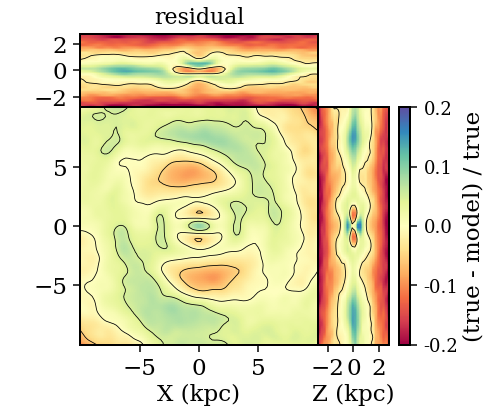}}\\
	\caption[]{  Comparison of the model inferred density distribution of mock galaxy $ I_1 $ with the simulation. (a) Surface density distributions of the simulation projected on $ x-y $, $ x-z $, and $ y-z $ planes. (b) Surface density distributions of a model with ($ \theta=58^{\circ}$, \hspace{0.05cm} $\varphi_{\mathrm{bar}}=-47^{\circ}$, \hspace{0.05cm} $ \psi_{\mathrm{bar}}=92^{\circ} $). Dashed lines in both panels mark the full length of the bar in  simulation ($ \sim 8 \hspace{0.1cm} \mathrm{kpc} $). (c) The residuals of (\subref{fig:dent1}) and (\subref{fig:dend}).
	}%
	\label{fig:ex3}%
\end{figure*}

We compare the model inferred density distribution with the true density distribution of the simulation.  In Fig.  \ref{fig:dent1}, we show the surface density distribution of the simulation projected on the $ x-y $, $ x-z $, and $ y-z $ planes. In Fig. \ref{fig:dend}, we show one model chosen from the allowed regions of viewing angles with $ (\theta , \varphi_{\mathrm{bar}}, \psi_{\mathrm{bar}}) = (58^{\circ}, -47^{\circ}, 92^{\circ}) $. We emphasize that $\theta$ is the same for the disk and the barred bulge as we align the barred bulge in the disk plane. Any $ \varphi $ is allowed for disk because it is assumed to be an oblate axisymmetric structure. We always have $ \psi_\mathrm{disk} = 90^{\circ} $ because the disk major axis is aligned with the $ x^{\prime} $-axis in the observational plane. The residual between the true and model inferred density distribution is shown in Fig. \ref{fig:res}.
\par The 3D density distribution from our model generally matches the original simulation, with a triaxial barred bulge located at the center of a disk. We obtain the bulge-dominated area along each principal axis $(x_\mathrm{bulge},  y_\mathrm{bulge}, z_\mathrm{bulge})$  as $(\sim 4 \hspace{0.1cm} \mathrm{kpc}, \sim 2 \hspace{0.1cm} \mathrm{kpc}, \sim 0.78 \hspace{0.1cm} \mathrm{kpc})$ for the mock galaxy $ I_{1} $, and $(\sim 3.85 \hspace{0.1cm} \mathrm{kpc}, \sim 2 \hspace{0.1cm} \mathrm{kpc}, \sim 0.65 \hspace{0.1cm} \mathrm{kpc})$ for our model inferred density (See appendix \ref{ap4} for details ). Our model does not have spiral structure. And the disk in our model is not thin enough in the outer regions, which is a common issue for disk deprojections if not seen perfectly edge-on. Our model also does not match the peanut shape of the barred bulge seen edge-on.  
\par The resulting density distributions for other mock galaxies ($ I_{2} $ to $ I_{6} $) are shown in appendix \ref{ap1}. 
The deprojection of disks are hard for galaxies with low inclination angles due to lack of information about the disk's intrinsic shapes, thus we do not try the method for galaxies with $\theta_{\mathrm{disk}} \lesssim 45^{\circ}$, while it is easier for edge-on galaxies. In contrast, the deprojection of barred bulges prefer galaxies with lower inclination angles, in which the shapes of the barred bulges are better revealed.
\par Overall, our method works reasonably well to match the basic shapes of disk and barred bulge for galaxies with moderate inclination angles and with different bar orientations, although the intrinsic shape of the barred bulge is harder to find when it is projected nearly end-on.

  \section{Verification of the deprojected model} \label{m-pot}
Before this model-inferred 3D density distribution could be used to build the gravitational potential of a dynamical model, we have to figure out how much uncertainty/bias the model might introduce. We use AGAMA \footnote{\url{https://github.com/GalacticDynamics-Oxford/Agama}} \cite[]{Vasiliev.2019a} to calculate the potential, forces, and orbits with our model inferred 3D density and then compare to those calculated with the true density distribution of the simulation. We illustrate the results with mock galaxy $I_{1}$.  

\subsection{Potential} \label{s-pot}
\begin{figure*}
	\centering
	\subfigure[][]{%
		\label{potf-1}%
		\includegraphics[width=5.95cm]{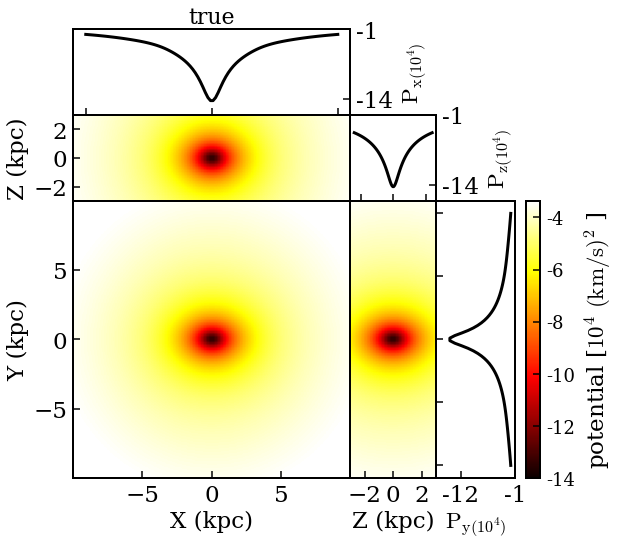}}%
	\hspace{8pt}%
	\subfigure[][]{%
		\label{potf-2}%
		\includegraphics[width=6cm]{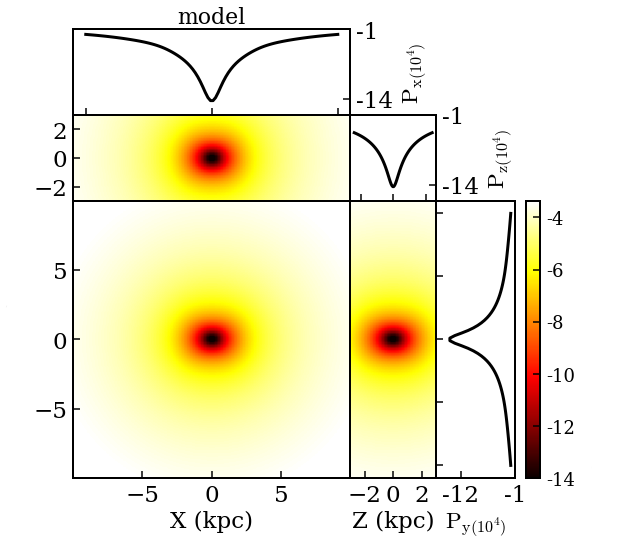}}
	\subfigure[][]{%
		\label{potf-4}%
		\includegraphics[width=5.4cm]{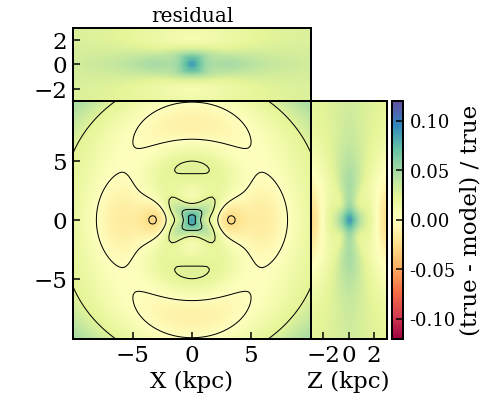}}\\
	\caption[]{Comparison of potentials of (\subref{potf-1}) the simulation in 2D planes and along the principle axes (solid black lines) and (\subref{potf-2}) our deprojected 3D model.
		(\subref{potf-4}) The residuals of (\subref{potf-1}) and (\subref{potf-4}).}%
	\label{pott}%
	
\end{figure*}

To obtain the gravitational potential given a density distribution, a numerical integration is required to solve the Poisson equation.
\begin{equation}
\label{poisson}
\nabla^{2} \Phi(\boldsymbol{x})=4 \pi G \rho(\boldsymbol{x}).
\end{equation}
 We freeze the N-body system at the snapshot that is chosen and calculate potential from the particle distribution using multipole expansion of spherical harmonics \cite[]{Binney.2008}. A similar method is adopted to compute the potential of model-inferred 3D density distribution given by Fig. \ref{fig:dend}.  The true and model potentials in 2D planes of $ x-y $, $ x-z $, $ y-z $, and along each principal axis are presented in Fig. \ref{potf-1} and  \ref{potf-2} respectively. The relative difference of the model and true potentials is less than $ 10 \% $ in all regions. 

\subsection{Force} \label{s-forc}
\begin{figure*}
	\centering	
	\subfigure[][]{%
		\label{fotf-1}%
		\includegraphics[width=5.95cm]{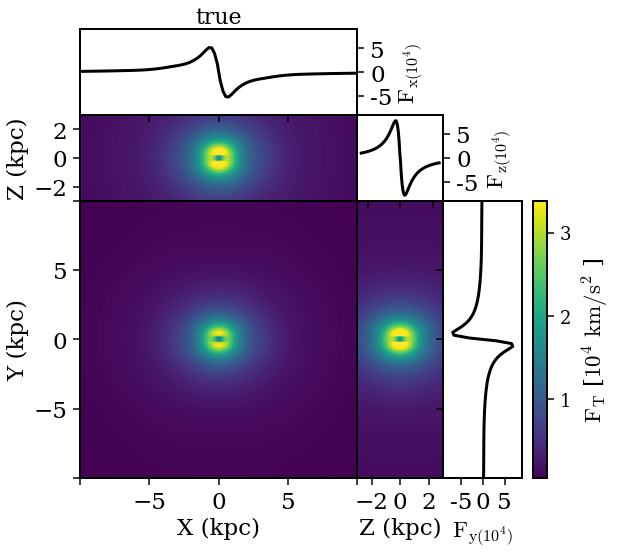}}%
	\hspace{8pt}%
	\subfigure[][]{%
		\label{fotf-2}%
		\includegraphics[width=6cm]{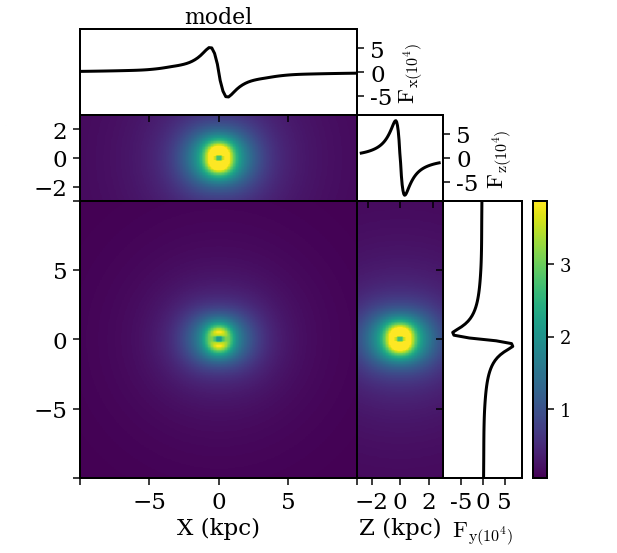}}
	\subfigure[][]{%
		\label{fotf-4}%
		\includegraphics[width=5.4cm]{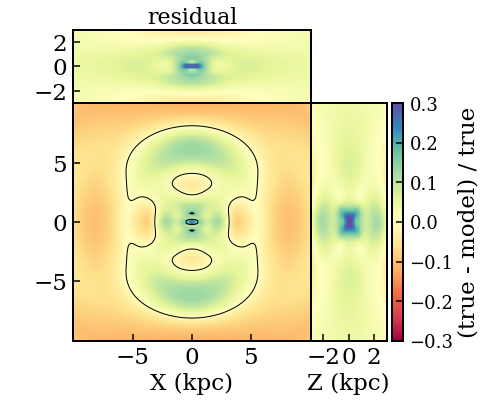}}\\

	\caption[]{Force comparison of (\subref{fotf-1}) the simulation in 2D planes and along principle axes (solid black lines) and (\subref{fotf-2}) our deprojected 3D model.  
		(\subref{fotf-4})  The residuals of (\subref{fotf-1}) and (\subref{fotf-2}).}%
	\label{fot}%
\end{figure*}

	The force is the derivative of potential at each position.
	\begin{equation}
	\label{for}
	\vec{F} = -\nabla \Phi, \hspace{0.5cm} |F_{T}|=\sqrt{F_{x}^{2}+F_{y}^{2}+F_{z}^{2}}.
	\end{equation}
	  Fig. \ref{fotf-1} and \ref{fotf-2} show the total force in the planes of $ x-y $, $ x-z $, $ y-z $, and force components along each axis for the true and model derived densities respectively.  The relative difference is less than $ 15 \% $ in all components and the maximum error occurs around the central point with $ 20 \% $.

\subsection{Orbital analysis} \label{s-orb}
 We further check if the potential derived from our model can support the key orbital families of the bar. A rigid pseudo-isothermal DM halo potential (as described in \S \ref{s-moc}) is added to the true and model potential, respectively. The scale velocity and scale radius are adopted as $V_{\mathrm{c}}=250  \hspace{.05cm} \mathrm{km}\mathrm{s}^{-1}$ and $R_{c}=15 \hspace{.05cm} \mathrm{kpc}$ respectively. We randomly select $ 15,000 $  initial conditions corresponding to the positions and velocities of particles from the snapshot at $ \mathrm{t} = 2.4 \hspace{.05cm} \mathrm{Gyr} $. We choose particles inside the corotation radius ($ R < 4.7 $ $ \mathrm{kpc} $),  then integrate the orbits in our model inferred potential and in the true potential.
\par AGAMA uses a modified version of the $ 8 $th order Runge-Kutta integrator $ \mathrm{DOP853} $ \cite[]{Hai.2008}. We calculate the orbits for the chosen initial conditions in a rotating frame. So the appropriate Coriolis and centrifugal pseudo-forces are taken into account and determined by the bar pattern speed $ \Omega_{ \mathrm{p}} $. We adopt $ \Omega_{ \mathrm{p}}=39 \hspace{.08cm} \mathrm{km \hspace{.04cm} s^{-1} \hspace{.04cm} kpc^{-1} }$, which is the true pattern speed of the simulated bar \cite[]{Shen.2014}. 

\subsubsection{Typical orbit families} \label{orb-test}
\par First, we visually check if orbits with the same initial conditions are similar in the true and model potentials. Here, orbits are integrated for $ 2.5 $ $ \mathrm{Gyr} $. In the true potential, the time period of a circular orbit at the end of the bar region is around  $ 0.1 $ $ \mathrm{Gyr} $. Among the $ 15,000 $ selected orbits, we randomly plot $ 200 $ orbits with the same initial conditions in both potentials.  Then we check the similarity in appearance of the true and model orbits in $ x-y $, $ x-z $, and $ y-z $ planes. Orbits with similar trajectories are considered as the matched ones.  We repeat this process a few more times and average the percentage of matched orbits in each selection. Generally, we conclude that $ 85 \%$  of all orbits in our model sample are visually matched with those in the true potential and around $ 15 \%$ are unmatched.
\begin{figure*}%
	\centering
	\subfigure[]{ %
		\label{fig:1orb}%
		\includegraphics[width=6.9cm]{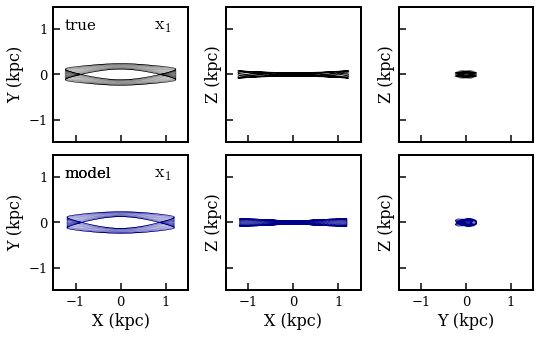}}    
	\subfigure[][]{ 
		\label{fig:2orb} 
		\includegraphics[width=7.1cm]{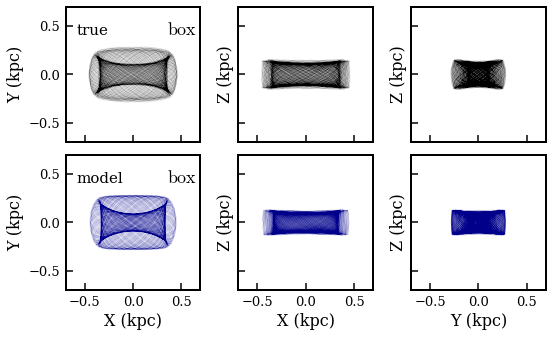}}
	\subfigure[][]{ 
		\label{fig:3orb} 
		\includegraphics[width=6.95cm]{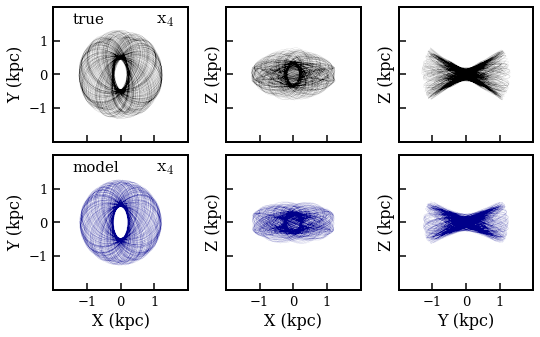}}%
	\subfigure[][]{ 
		\label{fig:4orb} 
		\includegraphics[width=7.1cm]{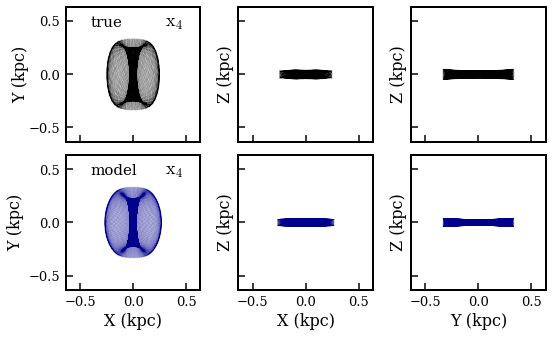}}
	\subfigure[][]{ 
		\label{fig:5orb} 
		\includegraphics[width=7cm]{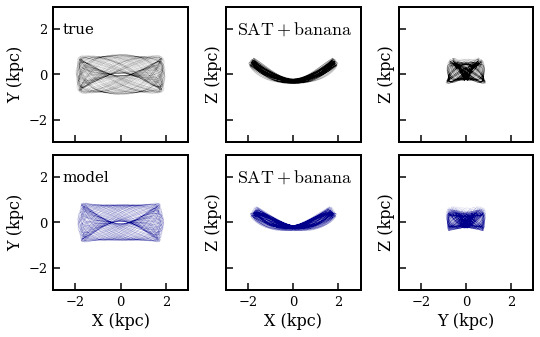}}
	\subfigure[][]{ 
		\label{fig:6orb} 
		\includegraphics[width=7cm]{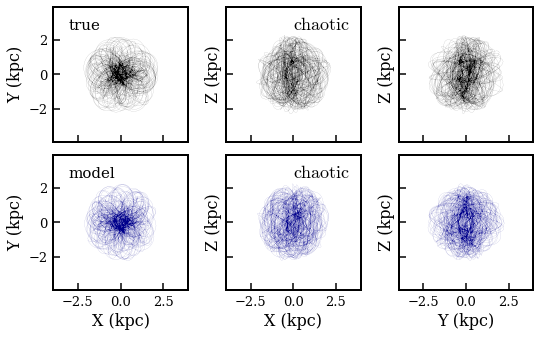}}
	\caption[]{Typical matched orbits in the true potential (black), and in our model (blue), plotted in the $ x-y $, $ x-z $, and $ y-z $ . Each pair of orbits are calculated with the same initial conditions.}%
	\label{fig:match}%
\end{figure*}

\par  The most important bar supporting orbit families are prograde $ x_1 $ family, which are elongated along the major axis of the bar. $ x_1 $ orbits in bars originate from the same parent as the box orbit family in triaxial potentials \cite[]{Contopoulos.1980, Schwarzschild.19822, Martinet.1988, Valluri.2016}. Other periodic orbit families in the bars are prograde $ x_2 $ and unstable $ x_3 $ orbits, which are elongated perpendicular to the bar and primarily found at small radii. In addition, retrograde $ x_4 $ orbits are also perpendicular to the bar and become nearly round at larger radii.
\par Some typical cases of matched orbits are shown in Fig. \ref{fig:match}.  The orbits calculated in the true simulated and our model potential are shown in black and blue respectively. The orbits in Fig. \ref{fig:1orb} and \ref{fig:2orb} are elongated along the bar major axis. They are $ x_1 $ and box orbit parented by $ x_1 $ orbits, respectively. While the orbits in Fig. \ref{fig:3orb} and \ref{fig:4orb} are retrograde $x_{4} $ orbits that are elongated perpendicular to the bar. In true and model potentials we did not find prograde $x_2$ and unstable $x_3 $ families in the inner region of the bar, similar to  \cite{Valluri.2016}. 
\par There are a few types of resonant orbits usually found in the N-body bar models: the orbits with $(\Omega_{x}, \Omega_{y}, \Omega_{z}) = (3:-2:0)$ called 'fish/pretzel' \cite[]{Valluri.2016}, the orbits with $(\Omega_{x},\Omega_{z}) = (1:2)$ known as 'banana' \cite[]{Pfenniger.1991} and the orbits with $(\Omega_{x}, \Omega_{y}, \Omega_{z}) = (3:0:-5)$ called 'brezel' \cite[]{2015.Portail}. The latter two types are proposed as the  backbone of X-shaped structure \cite[]{Patsis.2002, 2015.Portail}.
\par We find a few resonant orbits (< 2 \%) in the bar regions of our simulation. Fig. \ref{fig:5orb} shows a banana orbit in $ x-z $ plane that is well-matched in the true and model potentials. Some other types of resonant orbits can also be found in our model, but they could be originated from a different starting point than that similar orbit in the real potential (See appendix \ref{ap2}). Some of the apparently chaotic orbits can sweep similar regions in the true and model potential, as shown in Fig. \ref{fig:6orb}.
\begin{figure*}%
	\centering
	\includegraphics[width=18cm]{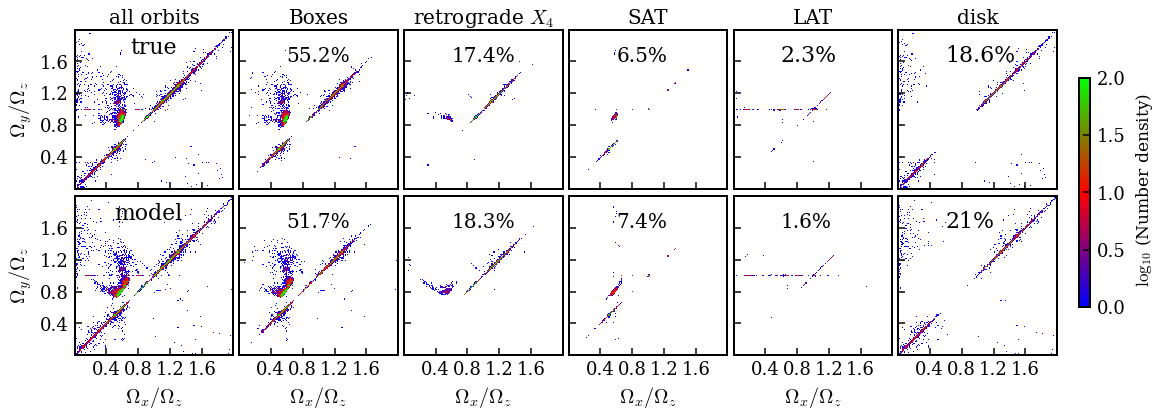}%
	\caption[]{Cartesian frequency map colored by number density, for orbits that are integrated for $ 13 $ $ \mathrm{Gyr} $ in true potential (top row) and our model (bottom row). Columns from left to right: frequency map for all orbits, boxes, retrograde $ x_{4} $, short-axis tubes (SAT) excluding retrograde $ x_{4} $, long-axis tubes (LAT), and disk orbits. The total number of all orbits is $ 15,000 $. }%
	\label{freqo}%
\end{figure*}

\subsubsection{Frequency analysis and orbit classification} \label{orb-test2}
\par Frequency analysis of orbits was first introduced by \cite{Binney.1982a, Binney.1984b} and later developed by \cite{Laskar.1990a, Laskar.1993b}. It is a powerful way to understand the features of orbits in large samples.  The fundamental orbital frequencies are obtained via Fourier transform of their position and velocity coordinates.  A long time integration for orbits is required to get an accurate frequency map (e.g., $ 20-50 $ orbital periods) \cite[]{Valluri.1998, Valluri.2016}. We integrate the orbits in our samples for $13$ $ \mathrm{Gyr} $. Then we use the NAFF software \footnote{\url{https://bitbucket.org/cjantonelli/naffrepo/src/master/}}  \cite[]{Valluri.1998, Valluri.2016} to compute the fundamental frequencies and perform automated bar orbit classification. Our computation is done in the Cartesian coordinates, which enables better classification of the bar orbits than in the cylindrical coordinates (see appendix B in \cite{Valluri.2016}).
\begin{figure*}
	\centering	%
	\includegraphics[width=18cm]{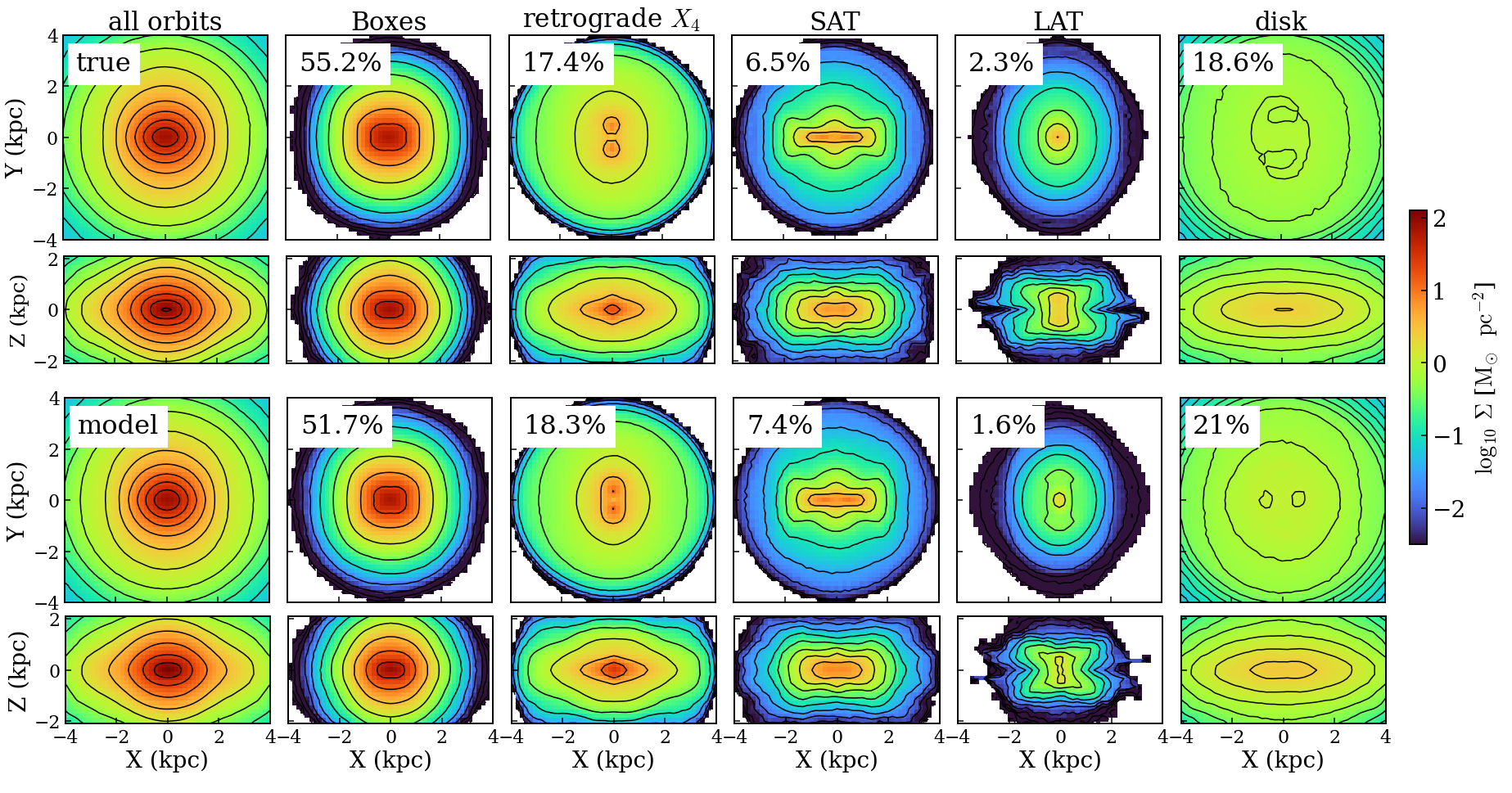}
	\hspace{8pt}%
	\caption[]{Projected surface density in $ x-y $ and $ x-z $ plane extracted from the 15,000 selected orbits in true potential (top rows) and model potential (bottom rows). Columns from left to right: surface densities for all orbits, boxes, retrograde $ x_{4} $, short-axis tubes (SAT) excluding retrograde $ x_{4} $, long-axis tubes (LAT), and disk orbits.}%
	\label{orbden}%
\end{figure*}
\par Frequency maps of our samples are shown in Fig. \ref{freqo}. The top and bottom rows are true and model orbits, respectively. The first column from left to right represent the frequency maps of all the 15,000 orbits in our sample. The frequency map of orbits in our model has generally the same features as that of orbits in the true potential, except there is a small frequency offset for the peak number density. The peak number density in frequency map of orbits in the true potential is around $ \Omega_{x} / \Omega_{z} = 0.65 $, $ \Omega_{y} / \Omega_{z} = 0.87 $ and for orbits in our model potential is around $ \Omega_{x} / \Omega_{z} = 0.60 $, $ \Omega_{y} / \Omega_{z} = 0.80 $.
\par We classify the orbits in our samples into boxes, retrograde $ x_{4} $, short-axis tubes (SAT), long-axis tubes (LAT), and disk orbits. We emphasize here SAT refer to all z-tube orbits excluding the retrograde $ x_{4} $. The frequency maps of each class of orbits are shown in Fig. \ref{freqo} from the second to the fifth columns (left to right), respectively.  Comparison of fractions of different classes of orbits in our model and in the true potential are shown in Table \ref{table:clas}. Box orbits contribute more than $ 50 $ percent of the orbits, while retrograde $ x_{4} $ orbits contribute $\sim 18$ percent. The SAT and LAT orbits have lower fractions. The orbital fractions in the true potential and in our model are generally consistent with each other.  We use ($ \mathrm{R_{apo}} > 4 \hspace{0.1cm} \mathrm{kpc} $) to determine the disk orbits, which means the orbits with apocenter radii larger than half-length of the bar ($ 4 \hspace{0.1cm}  \mathrm{kpc} $). The contributions of disk orbits are $18.6\% $ in the true potential and $21\% $ in our model potential.
\par We use the frequency drift parameter to determine the chaotic orbits \cite[]{Valluri2.2010}. In this method, the orbital time series are divided into two equal parts and the orbital fundamental frequencies are computed for each time segment. Since regular orbits have time-independent frequencies, the change in the frequency measured in the two time segments can be used to determine the diffusion rate in frequency space \cite[]{Laskar.1993, Valluri.1998}. The frequency drift for each frequency component $ \Omega_{i} $ ($ i $ = $ x $, $ y $, $ z $) can be computed as \cite[]{Valluri2.2010}:
\begin{equation}\log_{10} \left(\Delta f_{i}\right)=\log_{10} \left|\frac{\Omega_{i}\left(t_{1}\right)-\Omega_{i}\left(t_{2}\right)}{\Omega_{i}\left(t_{1}\right)}\right|.
\end{equation}
The frequency drift parameter $ \log_{10}(\Delta f) $ is defined as the value associated with the fundamental frequency $ \Omega_{i} $ with the largest amplitude in the Fourier spectrum. A larger frequency drift parameter indicates the orbit to be more chaotic. To separate chaotic orbits from regular orbits in frozen N-body potentials, $\log_{10} (\Delta f) >-1.2$  is a good empirical choice as tested in \cite{Valluri2.2010}. This is also a good choice for our samples (see appendix \ref{ap3}).
\par Overall $16\%$ of all orbits in the true potential and $17.6\%$ in our model potential have chaotic features. The contributions of all types of orbits in our model potential are similar to these in the true potential.

\begin{table}
	\centering
	\footnotesize
	\begin{tabular}{p{0.35\linewidth}p{0.20\linewidth}p{0.20\linewidth}}
		\hline
		\toprule
		class/type                   & true        & model     \\
		\midrule
		boxes                         & 55.2 \%    & 51.7\%    \\
		retrograde $ x_{4} $          & 17.4 \%    & 18.3\%    \\
		SAT                           & 6.5  \%    & 7.4\%     \\				
		LAT                           & 2.3  \%    & 1.6\%     \\
		disk orbits                   & 18.6 \%    & 21\%      \\
	    \midrule
	    sum  						  & 100 \%      & 100 \%    \\
		\midrule
		chaotic                       & 16 \%     & 17.6 \%    \\
		regular                       & 84 \%     & 82.4 \%    \\
		\bottomrule
	\end{tabular}
	\\
	\parbox{\columnwidth}{\caption{Classification of $ 15,000 $ selected orbits in the true and our model potentials. Orbits are classified into boxes, retrograde $ x_{4} $, short-axis tubes (SAT) excluding retrograde $ x_{4} $, long-axis tubes (LAT) and disk orbits. The bottom part of the table presents the fraction of orbits that have chaotic features in general.}
		\label{table:clas}}
\end{table}

\subsubsection{Structures constructed with orbits} \label{orb-test3}
To check if the structures could be built with the orbits in our model, we store each orbit in equal-time steps and sum up the density of the 15,000 selected orbits. 
\par The top rows in Fig. \ref{orbden} show surface densities in $ x-z $ and $ x-z $ planes constructed with orbits in the true potential. The bottom rows are with the orbits in the model potential. Columns from left to right are surface densities built by all orbit, boxes, retrograde $ x_{4} $, SAT, LAT, and disk orbits, respectively. In general, the structures generated by our model orbits are similar to the real ones. The retrograde $ x_{4} $  orbits support a structure perpendicular to the bar in face-on view, while the SAT orbits support a structure elongated along the bar in the face-on map and a boxy/peanut-shaped structure in the edge-on map. The LAT orbits support a structure elongated perpendicular to the bar in the face-on map and an X-shaped structure in the edge-on map. As we discussed in \S \ref{m-de}, the boxy/peanut-shaped structure is missing in our model inferred density distribution, however, as shown in the bottom panel of Fig. \ref{orbden},  the combination of SAT and LAT orbits in our model potential could still support such a boxy/peanut-shaped structure in the $ x-z $ plane.
\par In summary, we find that our model potential can support the major orbits families in a boxy/peanut-shaped bar. Although the fractions of various types of orbits in our model potential slightly differ from those in the true potential, we still have all types of orbits that could build the backbone of the bar.

\section{Conclusions} \label{s-con} 
We present a method of deprojecting 2D image of a barred galaxy to construct its 3D intrinsic density distribution. We adopt a two-step process, first, we decompose the galaxy into an elliptical, Sersic bulge, and a flat, exponential disk.  By subtracting the disk from the original image we get a barred bulge, then we fit the barred bulge and the disk separately with MGE and deproject the 2D MGE to get 3D intrinsic density distribution.
\par We assume the disk is axisymmetric, the barred bulge is triaxial, and the major axis of the barred bulge is aligned in the disk plane. Additional constraints are imposed to narrow down the parameter space of allowed viewing angles: (a) The position angle of the barred bulge is restricted to be close to that of the reference disk, subtracting the apparent difference $\Delta \psi^\prime$ measured from MGE fitting. (b) the inclination angle of the disk is restricted to be close to the value obtained by assuming an intrinsic flattening of $\mathrm{q}_{\mathrm{int}}=0.26$, (c) the intrinsic scale height of the outer barred bulge is assumed to be similar to the disk in the same region.
\par By combining the 3D density distribution of a barred bulge and a disk, we construct the 3D density distribution of the whole galaxy. We validate the method by applying it to mock images created from a simulated barred galaxy. By comparing with the true simulation, we find that: \\  1) In general, the 3D density distribution from our model matches the true simulation, with a triaxial barred bulge located at the center of a disk. However, our model does not match the boxy/peanut shape of the bulge when seen edge-on. Meanwhile, the disk in our model does not have spiral structure. And the disk is not very thin at the outer regions in our model, which is a common issue for disk deprojection if not seen edge-on. 
\\ 2) We verify this method by comparing the potential and force inferred from the model constructed 3D density to the true value of the simulation. The residuals of subtracting the model potential from the true one can be up to $ 10 \% $ in an extended region within the barred bulge. While the difference in forces can be as large as $ 15 \% $ in the barred bulge region and $ 20 \% $ around the very center. \\ 3) We find that 85$ \% $ of our sample orbits, including the major families of bar-supporting orbits, resonant and chaotic orbits, in the model potential turn out to be very similar to those in the true potential. The unmatched 15\% are mostly chaotic or resonant orbits, their morphologies are easily altered due to small differences in the gravitational field.
\\ 4) The orbits in the model inferred potential distribute similarly in the frequency maps as those in the true potential. We classify orbits into boxes, retrograde $ x_{4} $, short-axis tubes excluding retrograde $ x_{4} $, long-axis tubes, and disk orbits. The contributions of different classes of orbits in our model potential are close to these in the true potentials. The short-axis tubes in our model build an elongated bar with box/peanut-shaped structure, and the long-axis tubes build a X-shaped structure. These match perfectly the structures in the true potential.

\par We have shown that this method can construct 3D density of barred galaxies from their 2D images, Although our model inferred density does not match the boxy/peanut shape exactly, the potential still supports the major orbit families reproducing the boxy/peanut-shaped and X-shaped structures. The fractions of orbits that building the backbone of bar in our model slightly differ from those in the true potential. In a dynamical model, we have the freedom of giving different weights to different orbits by fitting the data. Thus the fraction of bar-supporting orbits in our model should not in principle matter once we have sampled all typical orbits. In the future, we will test to construct Schwarzschild/M2M models of nearby barred galaxies.

\section*{Acknowledgements}
We thank Eugene Vasiliev and Monica Valluri for useful discussions.
The research presented here is partially supported by the National Key R\&D Program of China under grant No. 2018YFA0404501; by the National Natural Science Foundation of China under grant Nos. 12025302, 11773052, 11761131016; by the ``111'' Project of the Ministry of Education of China under grant No. B20019; and by the Chinese Space Station Telescope project, and by the Deutsche Forschungsgemeinschaft under grant GZ GE 567/5-1(OG). This work made use of the Gravity Supercomputer at the Department of Astronomy, Shanghai Jiao Tong University, and the facilities of the Center for High Performance Computing at Shanghai Astronomical Observatory. LZ acknowledges the support from National Natural Science Foundation of China under grant No. Y945271001. BT acknowledges support from  CAS-TWAS President's Fellowship for international PhD students, awarded jointly by the Chinese Academy of science and The World Academy of Sciences.
\\
\\
\textit{Software}: \textit{Agama} \cite[]{Vasiliev.2019a}, \textit{Astropy} \cite[]{astropy.2013, astropy.2018}, \textit{Galfit} \cite[]{Peng.2010}, \textit{Jupyter Notebook} \cite[]{Kluyver.2016}, \textit{MgeFit} \cite[]{cappellari.2002},  \textit{matplotlib} \cite[]{Hunter.2007}, \textit{numpy} \cite[]{Harris.2020}, \textit{photutils} \cite[]{Bradley.2016}, \textit{NAFF} \cite[]{Valluri.1998, Valluri.2016}, \textit{scipy} \cite[]{Virtanen.2020}.  
\section*{DATA AVAILABILITY}
The data underlying this article will be shared on reasonable request to the corresponding author.

	
	
\bibliographystyle{mnras}
\bibliography{library}

\begin{thebibliography}{}
\makeatletter
\relax
\def\mn@urlcharsother{\let\do\@makeother \do\$\do\&\do\#\do\^\do\_\do\%\do\~}
\def\mn@doi{\begingroup\mn@urlcharsother \@ifnextchar [ {\mn@doi@}
  {\mn@doi@[]}}
\def\mn@doi@[#1]#2{\def\@tempa{#1}\ifx\@tempa\@empty \href
  {http://dx.doi.org/#2} {doi:#2}\else \href {http://dx.doi.org/#2} {#1}\fi
  \endgroup}
\def\mn@eprint#1#2{\mn@eprint@#1:#2::\@nil}
\def\mn@eprint@arXiv#1{\href {http://arxiv.org/abs/#1} {{\tt arXiv:#1}}}
\def\mn@eprint@dblp#1{\href {http://dblp.uni-trier.de/rec/bibtex/#1.xml}
  {dblp:#1}}
\def\mn@eprint@#1:#2:#3:#4\@nil{\def\@tempa {#1}\def\@tempb {#2}\def\@tempc
  {#3}\ifx \@tempc \@empty \let \@tempc \@tempb \let \@tempb \@tempa \fi \ifx
  \@tempb \@empty \def\@tempb {arXiv}\fi \@ifundefined
  {mn@eprint@\@tempb}{\@tempb:\@tempc}{\expandafter \expandafter \csname
  mn@eprint@\@tempb\endcsname \expandafter{\@tempc}}}

\bibitem[\protect\citeauthoryear{{Aguerri}, {M{\'e}ndez-Abreu}  \&
  {Corsini}}{{Aguerri} et~al.}{2009}]{Aguerri.2009}
{Aguerri} J.~A.~L.,  {M{\'e}ndez-Abreu} J.,   {Corsini} E.~M.,  2009, \mn@doi
  [\aap] {10.1051/0004-6361:200810931}, \href
  {https://ui.adsabs.harvard.edu/abs/2009A&A...495..491A} {495, 491}

\bibitem[\protect\citeauthoryear{{Astropy Collaboration} et~al.,}{{Astropy
  Collaboration} et~al.}{2013}]{astropy.2013}
{Astropy Collaboration} et~al., 2013, \mn@doi [\aap]
  {10.1051/0004-6361/201322068}, \href
  {https://ui.adsabs.harvard.edu/abs/2013A&A...558A..33A} {558, A33}

\bibitem[\protect\citeauthoryear{{Astropy Collaboration} et~al.,}{{Astropy
  Collaboration} et~al.}{2018}]{astropy.2018}
{Astropy Collaboration} et~al., 2018, \mn@doi [\aj] {10.3847/1538-3881/aabc4f},
  \href {https://ui.adsabs.harvard.edu/abs/2018AJ....156..123A} {156, 123}

\bibitem[\protect\citeauthoryear{{Athanassoula}}{{Athanassoula}}{2003}]{2003.Athanassoula}
{Athanassoula} E.,  2003, \mn@doi [\mnras] {10.1046/j.1365-8711.2003.06473.x},
  \href {https://ui.adsabs.harvard.edu/abs/2003MNRAS.341.1179A} {341, 1179}

\bibitem[\protect\citeauthoryear{{Barazza}, {Jogee}  \& {Marinova}}{{Barazza}
  et~al.}{2008}]{Barazza.2008}
{Barazza} F.~D.,  {Jogee} S.,   {Marinova} I.,  2008, \mn@doi [\apj]
  {10.1086/526510}, \href
  {https://ui.adsabs.harvard.edu/abs/2008ApJ...675.1194B} {675, 1194}

\bibitem[\protect\citeauthoryear{{Bendinelli}}{{Bendinelli}}{1991}]{Bendinelli.1991}
{Bendinelli} O.,  1991, \mn@doi [\apj] {10.1086/169595}, \href
  {https://ui.adsabs.harvard.edu/abs/1991ApJ...366..599B} {366, 599}

\bibitem[\protect\citeauthoryear{{Binney}}{{Binney}}{1985}]{binny.1985}
{Binney} J.,  1985, \mn@doi [\mnras] {10.1093/mnras/212.4.767}, \href
  {https://ui.adsabs.harvard.edu/abs/1985MNRAS.212..767B} {212, 767}

\bibitem[\protect\citeauthoryear{{Binney} \& {Spergel}}{{Binney} \&
  {Spergel}}{1982}]{Binney.1982a}
{Binney} J.,  {Spergel} D.,  1982, \mn@doi [\apj] {10.1086/159559}, \href
  {https://ui.adsabs.harvard.edu/abs/1982ApJ...252..308B} {252, 308}

\bibitem[\protect\citeauthoryear{{Binney} \& {Spergel}}{{Binney} \&
  {Spergel}}{1984}]{Binney.1984b}
{Binney} J.,  {Spergel} D.,  1984, \mn@doi [\mnras] {10.1093/mnras/206.1.159},
  \href {https://ui.adsabs.harvard.edu/abs/1984MNRAS.206..159B} {206, 159}

\bibitem[\protect\citeauthoryear{{Binney} \& {Tremaine}}{{Binney} \&
  {Tremaine}}{2008}]{Binney.2008}
{Binney} J.,  {Tremaine} S.,  2008, {Galactic Dynamics: Second Edition}

\bibitem[\protect\citeauthoryear{{Bissantz} \& {Gerhard}}{{Bissantz} \&
  {Gerhard}}{2002}]{Gerhard.2002}
{Bissantz} N.,  {Gerhard} O.,  2002, \mn@doi [\mnras]
  {10.1046/j.1365-8711.2002.05116.x}, \href
  {https://doi.org/10.1046/j.1365-8711.2002.05116.x} {330, 591}

\bibitem[\protect\citeauthoryear{{Bla{\~n}a D{\'\i}az} et~al.,}{{Bla{\~n}a
  D{\'\i}az} et~al.}{2018}]{Bla.2018}
{Bla{\~n}a D{\'\i}az} M.,  et~al., 2018, \mn@doi [\mnras]
  {10.1093/mnras/sty2311}, \href
  {https://ui.adsabs.harvard.edu/abs/2018MNRAS.481.3210B} {481, 3210}

\bibitem[\protect\citeauthoryear{{Bradley} et~al.,}{{Bradley}
  et~al.}{2016}]{Bradley.2016}
{Bradley} L.,  et~al., 2016, {Photutils: Photometry tools} (\mn@eprint {ascl}
  {1609.011})

\bibitem[\protect\citeauthoryear{{Breddels}, {Helmi}, {van den Bosch}, {van de
  Ven}  \& {Battaglia}}{{Breddels} et~al.}{2013}]{2013.Breddels}
{Breddels} M.~A.,  {Helmi} A.,  {van den Bosch} R.~C.~E.,  {van de Ven} G.,
  {Battaglia} G.,  2013, \mn@doi [\mnras] {10.1093/mnras/stt956}, \href
  {https://ui.adsabs.harvard.edu/abs/2013MNRAS.433.3173B} {433, 3173}

\bibitem[\protect\citeauthoryear{{Brown}, {Valluri}, {Shen}  \&
  {Debattista}}{{Brown} et~al.}{2013}]{Brown.2013}
{Brown} J.~S.,  {Valluri} M.,  {Shen} J.,   {Debattista} V.~P.,  2013, \mn@doi
  [\apj] {10.1088/0004-637X/778/2/151}, \href
  {https://ui.adsabs.harvard.edu/abs/2013ApJ...778..151B} {778, 151}

\bibitem[\protect\citeauthoryear{{Bureau} \& {Athanassoula}}{{Bureau} \&
  {Athanassoula}}{2005}]{2005.Bureau}
{Bureau} M.,  {Athanassoula} E.,  2005, \mn@doi [\apj] {10.1086/430056}, \href
  {https://ui.adsabs.harvard.edu/abs/2005ApJ...626..159B} {626, 159}

\bibitem[\protect\citeauthoryear{{Cappellari}}{{Cappellari}}{2002}]{cappellari.2002}
{Cappellari} M.,  2002, \mn@doi [\mnras] {10.1046/j.1365-8711.2002.05412.x},
  \href {https://ui.adsabs.harvard.edu/abs/2002MNRAS.333..400C} {333, 400}

\bibitem[\protect\citeauthoryear{{Cappellari}}{{Cappellari}}{2008}]{Cappellari.2008}
{Cappellari} M.,  2008, \mn@doi [\mnras] {10.1111/j.1365-2966.2008.13754.x},
  \href {https://ui.adsabs.harvard.edu/abs/2008MNRAS.390...71C} {390, 71}

\bibitem[\protect\citeauthoryear{{Cappellari} et~al.,}{{Cappellari}
  et~al.}{2006}]{Cappellari.2006}
{Cappellari} M.,  et~al., 2006, \mn@doi [\mnras]
  {10.1111/j.1365-2966.2005.09981.x}, \href
  {https://ui.adsabs.harvard.edu/abs/2006MNRAS.366.1126C} {366, 1126}

\bibitem[\protect\citeauthoryear{{Contopoulos}}{{Contopoulos}}{1956}]{Contopoulos.1956}
{Contopoulos} G.,  1956, \zap, \href
  {https://ui.adsabs.harvard.edu/abs/1956ZA.....39..126C} {39, 126}

\bibitem[\protect\citeauthoryear{{Contopoulos}}{{Contopoulos}}{1980}]{Contopoulos.1980}
{Contopoulos} G.,  1980, \aap, \href
  {https://ui.adsabs.harvard.edu/abs/1980A&A....81..198C} {81, 198}

\bibitem[\protect\citeauthoryear{{Cretton} \& {van den Bosch}}{{Cretton} \&
  {van den Bosch}}{1999}]{Cretton.1999}
{Cretton} N.,  {van den Bosch} F.~C.,  1999, \mn@doi [\apj] {10.1086/306971},
  \href {https://ui.adsabs.harvard.edu/abs/1999ApJ...514..704C} {514, 704}

\bibitem[\protect\citeauthoryear{{Cretton}, {de Zeeuw}, {van der Marel}  \&
  {Rix}}{{Cretton} et~al.}{1999}]{1999.Cretton}
{Cretton} N.,  {de Zeeuw} P.~T.,  {van der Marel} R.~P.,   {Rix} H.-W.,  1999,
  \mn@doi [\apjs] {10.1086/313264}, \href
  {https://ui.adsabs.harvard.edu/abs/1999ApJS..124..383C} {124, 383}

\bibitem[\protect\citeauthoryear{{Debattista} \& {Sellwood}}{{Debattista} \&
  {Sellwood}}{1998}]{1998.Debattista}
{Debattista} V.~P.,  {Sellwood} J.~A.,  1998, \mn@doi [\apjl] {10.1086/311118},
  \href {https://ui.adsabs.harvard.edu/abs/1998ApJ...493L...5D} {493, L5}

\bibitem[\protect\citeauthoryear{{Emsellem}, {Monnet}  \& {Bacon}}{{Emsellem}
  et~al.}{1994a}]{Emsellema.1994}
{Emsellem} E.,  {Monnet} G.,   {Bacon} R.,  1994a, \aap, \href
  {https://ui.adsabs.harvard.edu/abs/1994A%26A...285..723E} {285, 723}

\bibitem[\protect\citeauthoryear{{Emsellem}, {Monnet}, {Bacon}  \&
  {Nieto}}{{Emsellem} et~al.}{1994b}]{Emsellemb.1994}
{Emsellem} E.,  {Monnet} G.,  {Bacon} R.,   {Nieto} J.-L.,  1994b, \aap, \href
  {https://ui.adsabs.harvard.edu/abs/1994A%26A...285..739E} {285, 739}

\bibitem[\protect\citeauthoryear{{Erwin}}{{Erwin}}{2018}]{Erwin.2018}
{Erwin} P.,  2018, \mn@doi [\mnras] {10.1093/mnras/stx3117}, \href
  {https://ui.adsabs.harvard.edu/abs/2018MNRAS.474.5372E} {474, 5372}

\bibitem[\protect\citeauthoryear{{Erwin} \& {Debattista}}{{Erwin} \&
  {Debattista}}{2016}]{Erwin.2016}
{Erwin} P.,  {Debattista} V.~P.,  2016, \mn@doi [\apjl]
  {10.3847/2041-8205/825/2/L30}, \href
  {https://ui.adsabs.harvard.edu/abs/2016ApJ...825L..30E} {825, L30}

\bibitem[\protect\citeauthoryear{{Eskridge} et~al.,}{{Eskridge}
  et~al.}{2000}]{Eskridge.2000}
{Eskridge} P.~B.,  et~al., 2000, \mn@doi [\aj] {10.1086/301203}, 119, 536

\bibitem[\protect\citeauthoryear{{Fragkoudi}, {Athanassoula}, {Bosma}  \&
  {Iannuzzi}}{{Fragkoudi} et~al.}{2015}]{Fragkoudi.2015}
{Fragkoudi} F.,  {Athanassoula} E.,  {Bosma} A.,   {Iannuzzi} F.,  2015,
  \mn@doi [\mnras] {10.1093/mnras/stv537}, \href
  {https://ui.adsabs.harvard.edu/abs/2015MNRAS.450..229F} {450, 229}

\bibitem[\protect\citeauthoryear{{Friedli} \& {Benz}}{{Friedli} \&
  {Benz}}{1993}]{1993.Friedli}
{Friedli} D.,  {Benz} W.,  1993, \aap, \href
  {https://ui.adsabs.harvard.edu/abs/1993A&A...268...65F} {268, 65}

\bibitem[\protect\citeauthoryear{{Gadotti}}{{Gadotti}}{2009}]{Gaddoti.2009}
{Gadotti} D.~A.,  2009, \mn@doi [\mnras] {10.1111/j.1365-2966.2008.14257.x},
  \href {https://ui.adsabs.harvard.edu/abs/2009MNRAS.393.1531G} {393, 1531}

\bibitem[\protect\citeauthoryear{{Gadotti}}{{Gadotti}}{2011}]{2011.Gadotti}
{Gadotti} D.~A.,  2011, \mn@doi [\mnras] {10.1111/j.1365-2966.2011.18945.x},
  \href {https://ui.adsabs.harvard.edu/abs/2011MNRAS.415.3308G} {415, 3308}

\bibitem[\protect\citeauthoryear{{Gadotti}, {Athanassoula}, {Carrasco},
  {Bosma}, {de Souza}  \& {Recillas}}{{Gadotti} et~al.}{2007}]{Gadotti.2007}
{Gadotti} D.~A.,  {Athanassoula} E.,  {Carrasco} L.,  {Bosma} A.,  {de Souza}
  R.~E.,   {Recillas} E.,  2007, \mn@doi [\mnras]
  {10.1111/j.1365-2966.2007.12295.x}, \href
  {https://ui.adsabs.harvard.edu/abs/2007MNRAS.381..943G} {381, 943}

\bibitem[\protect\citeauthoryear{Gebhardt et~al.,}{Gebhardt
  et~al.}{2000}]{Gebhardt.2000}
Gebhardt K.,  et~al., 2000, \mn@doi [The Astronomical Journal]
  {10.1086/301240}, 119, 1157

\bibitem[\protect\citeauthoryear{{Hairer}, {Wanner}  \& {Norsett}}{{Hairer}
  et~al.}{1993}]{Hai.2008}
{Hairer} E.,  {Wanner} G.,   {Norsett} S.,  1993, {Solving Ordinary
  Differential Equations}.
Springer

\bibitem[\protect\citeauthoryear{{Harris} et~al.,}{{Harris}
  et~al.}{2020}]{Harris.2020}
{Harris} C.~R.,  et~al., 2020, \mn@doi [\nat] {10.1038/s41586-020-2649-2},
  \href {https://ui.adsabs.harvard.edu/abs/2020Natur.585..357H} {585, 357}

\bibitem[\protect\citeauthoryear{{Hunt}, {Kawata}  \& {Martel}}{{Hunt}
  et~al.}{2013}]{2013.Hunt}
{Hunt} J. A.~S.,  {Kawata} D.,   {Martel} H.,  2013, \mn@doi [\mnras]
  {10.1093/mnras/stt657}, \href
  {https://ui.adsabs.harvard.edu/abs/2013MNRAS.432.3062H} {432, 3062}

\bibitem[\protect\citeauthoryear{{Hunter}}{{Hunter}}{2007}]{Hunter.2007}
{Hunter} J.~D.,  2007, \mn@doi [Computing in Science and Engineering]
  {10.1109/MCSE.2007.55}, \href
  {https://ui.adsabs.harvard.edu/abs/2007CSE.....9...90H} {9, 90}

\bibitem[\protect\citeauthoryear{Kluyver et~al.,}{Kluyver
  et~al.}{2016}]{Kluyver.2016}
Kluyver T.,  et~al., 2016, in Loizides F.,  Schmidt B.,  eds, Positioning and
  Power in Academic Publishing: Players, Agents and Agendas. pp 87 -- 90

\bibitem[\protect\citeauthoryear{{Kormendy} \& {Kennicutt}}{{Kormendy} \&
  {Kennicutt}}{2004}]{KK.2004}
{Kormendy} J.,  {Kennicutt} Robert~C. J.,  2004, \mn@doi [\araa]
  {10.1146/annurev.astro.42.053102.134024}, \href
  {https://ui.adsabs.harvard.edu/abs/2004ARA&A..42..603K} {42, 603}

\bibitem[\protect\citeauthoryear{{Kowalczyk}, {{\L}okas}  \&
  {Valluri}}{{Kowalczyk} et~al.}{2017}]{2017.Kowalczyk}
{Kowalczyk} K.,  {{\L}okas} E.~L.,   {Valluri} M.,  2017, \mn@doi [\mnras]
  {10.1093/mnras/stx1520}, \href
  {https://ui.adsabs.harvard.edu/abs/2017MNRAS.470.3959K} {470, 3959}

\bibitem[\protect\citeauthoryear{{Laskar}}{{Laskar}}{1990}]{Laskar.1990a}
{Laskar} J.,  1990, \mn@doi [\icarus] {10.1016/0019-1035(90)90084-M}, \href
  {https://ui.adsabs.harvard.edu/abs/1990Icar...88..266L} {88, 266}

\bibitem[\protect\citeauthoryear{{Laskar}}{{Laskar}}{1993a}]{Laskar.1993b}
{Laskar} J.,  1993a, \mn@doi [Celestial Mechanics and Dynamical Astronomy]
  {10.1007/BF00699731}, \href
  {https://ui.adsabs.harvard.edu/abs/1993CeMDA..56..191L} {56, 191}

\bibitem[\protect\citeauthoryear{{Laskar}}{{Laskar}}{1993b}]{Laskar.1993}
{Laskar} J.,  1993b, \mn@doi [Physica D Nonlinear Phenomena]
  {10.1016/0167-2789(93)90210-R}, \href
  {https://ui.adsabs.harvard.edu/abs/1993PhyD...67..257L} {67, 257}

\bibitem[\protect\citeauthoryear{{Li} \& {Shen}}{{Li} \&
  {Shen}}{2012}]{li.2012}
{Li} Z.-Y.,  {Shen} J.,  2012, \mn@doi [\apjl] {10.1088/2041-8205/757/1/L7},
  \href {https://ui.adsabs.harvard.edu/abs/2012ApJ...757L...7L} {757, L7}

\bibitem[\protect\citeauthoryear{{Li}, {Ho}, {Barth}  \& {Peng}}{{Li}
  et~al.}{2011}]{LiYu.2011}
{Li} Z.-Y.,  {Ho} L.~C.,  {Barth} A.~J.,   {Peng} C.~Y.,  2011, \mn@doi [\apjs]
  {10.1088/0067-0049/197/2/22}, \href
  {https://ui.adsabs.harvard.edu/abs/2011ApJS..197...22L} {197, 22}

\bibitem[\protect\citeauthoryear{{Li}, {Shen}, {Bureau}, {Zhou}, {Du}  \&
  {Debattista}}{{Li} et~al.}{2018}]{2018.Li}
{Li} Z.-Y.,  {Shen} J.,  {Bureau} M.,  {Zhou} Y.,  {Du} M.,   {Debattista}
  V.~P.,  2018, \mn@doi [\apj] {10.3847/1538-4357/aaa771}, \href
  {https://ui.adsabs.harvard.edu/abs/2018ApJ...854...65L} {854, 65}

\bibitem[\protect\citeauthoryear{{Long}, {Mao}, {Shen}  \& {Wang}}{{Long}
  et~al.}{2013}]{2013.Long}
{Long} R.~J.,  {Mao} S.,  {Shen} J.,   {Wang} Y.,  2013, \mn@doi [\mnras]
  {10.1093/mnras/sts285}, \href
  {https://ui.adsabs.harvard.edu/abs/2013MNRAS.428.3478L} {428, 3478}

\bibitem[\protect\citeauthoryear{{Magorrian}}{{Magorrian}}{1999}]{Magorian.1999}
{Magorrian} J.,  1999, \mn@doi [\mnras] {10.1046/j.1365-8711.1999.02135.x},
  \href {https://ui.adsabs.harvard.edu/abs/1999MNRAS.302..530M} {302, 530}

\bibitem[\protect\citeauthoryear{{Martinet} \& {de Zeeuw}}{{Martinet} \& {de
  Zeeuw}}{1988}]{Martinet.1988}
{Martinet} L.,  {de Zeeuw} T.,  1988, \aap, \href
  {https://ui.adsabs.harvard.edu/abs/1988A&A...206..269M} {206, 269}

\bibitem[\protect\citeauthoryear{{Masters} et~al.,}{{Masters}
  et~al.}{2011}]{2011.Masters}
{Masters} K.~L.,  et~al., 2011, \mn@doi [\mnras]
  {10.1111/j.1365-2966.2010.17834.x}, \href
  {https://ui.adsabs.harvard.edu/abs/2011MNRAS.411.2026M} {411, 2026}

\bibitem[\protect\citeauthoryear{{M{\'e}ndez-Abreu}, {Simonneau}, {Aguerri}  \&
  {Corsini}}{{M{\'e}ndez-Abreu} et~al.}{2010}]{Abreu.2010}
{M{\'e}ndez-Abreu} J.,  {Simonneau} E.,  {Aguerri} J.~A.~L.,   {Corsini} E.~M.,
   2010, \mn@doi [\aap] {10.1051/0004-6361/201014130}, \href
  {https://ui.adsabs.harvard.edu/abs/2010A&A...521A..71M} {521, A71}

\bibitem[\protect\citeauthoryear{{M{\'e}ndez-Abreu}, {Costantin}, {Aguerri},
  {de Lorenzo-C{\'a}ceres}  \& {Corsini}}{{M{\'e}ndez-Abreu}
  et~al.}{2018}]{Abreu.2018}
{M{\'e}ndez-Abreu} J.,  {Costantin} L.,  {Aguerri} J.~A.~L.,  {de
  Lorenzo-C{\'a}ceres} A.,   {Corsini} E.~M.,  2018, \mn@doi [\mnras]
  {10.1093/mnras/sty1694}, \href
  {https://ui.adsabs.harvard.edu/abs/2018MNRAS.479.4172M} {479, 4172}

\bibitem[\protect\citeauthoryear{{Men{\'e}ndez-Delmestre}, {Sheth},
  {Schinnerer}, {Jarrett}  \& {Scoville}}{{Men{\'e}ndez-Delmestre}
  et~al.}{2007}]{Delmestre.2007}
{Men{\'e}ndez-Delmestre} K.,  {Sheth} K.,  {Schinnerer} E.,  {Jarrett} T.~H.,
  {Scoville} N.~Z.,  2007, \mn@doi [\apj] {10.1086/511025}, \href
  {https://ui.adsabs.harvard.edu/abs/2007ApJ...657..790M} {657, 790}

\bibitem[\protect\citeauthoryear{{Monnet}, {Bacon}  \& {Emsellem}}{{Monnet}
  et~al.}{1992}]{Emsellem.1992}
{Monnet} G.,  {Bacon} R.,   {Emsellem} E.,  1992, \aap, \href
  {https://ui.adsabs.harvard.edu/abs/1992A%26A...253..366M} {253, 366}

\bibitem[\protect\citeauthoryear{{Noordermeer} \& {van der
  Hulst}}{{Noordermeer} \& {van der Hulst}}{2007}]{Noordermeer.2007}
{Noordermeer} E.,  {van der Hulst} J.~M.,  2007, \mn@doi [\mnras]
  {10.1111/j.1365-2966.2007.11532.x}, \href
  {https://ui.adsabs.harvard.edu/abs/2007MNRAS.376.1480N} {376, 1480}

\bibitem[\protect\citeauthoryear{{Patsis}, {Skokos}  \&
  {Athanassoula}}{{Patsis} et~al.}{2002}]{Patsis.2002}
{Patsis} P.~A.,  {Skokos} C.,   {Athanassoula} E.,  2002, \mn@doi [\mnras]
  {10.1046/j.1365-8711.2002.05943.x}, \href
  {https://ui.adsabs.harvard.edu/abs/2002MNRAS.337..578P} {337, 578}

\bibitem[\protect\citeauthoryear{{Peng}, {Ho}, {Impey}  \& {Rix}}{{Peng}
  et~al.}{2010}]{Peng.2010}
{Peng} C.~Y.,  {Ho} L.~C.,  {Impey} C.~D.,   {Rix} H.-W.,  2010, \mn@doi [\aj]
  {10.1088/0004-6256/139/6/2097}, \href
  {https://ui.adsabs.harvard.edu/abs/2010AJ....139.2097P} {139, 2097}

\bibitem[\protect\citeauthoryear{{Pfenniger} \& {Friedli}}{{Pfenniger} \&
  {Friedli}}{1991}]{Pfenniger.1991}
{Pfenniger} D.,  {Friedli} D.,  1991, \aap, \href
  {https://ui.adsabs.harvard.edu/abs/1991A&A...252...75P} {252, 75}

\bibitem[\protect\citeauthoryear{{Portail}, {Wegg}, {Gerhard}  \&
  {Martinez-Valpuesta}}{{Portail} et~al.}{2015}]{2015.Portail}
{Portail} M.,  {Wegg} C.,  {Gerhard} O.,   {Martinez-Valpuesta} I.,  2015,
  \mn@doi [\mnras] {10.1093/mnras/stv058}, \href
  {https://ui.adsabs.harvard.edu/abs/2015MNRAS.448..713P} {448, 713}

\bibitem[\protect\citeauthoryear{{Portail}, {Gerhard}, {Wegg}  \&
  {Ness}}{{Portail} et~al.}{2017}]{Portail.2017}
{Portail} M.,  {Gerhard} O.,  {Wegg} C.,   {Ness} M.,  2017, \mn@doi [\mnras]
  {10.1093/mnras/stw2819}, \href
  {https://ui.adsabs.harvard.edu/abs/2017MNRAS.465.1621P} {465, 1621}

\bibitem[\protect\citeauthoryear{{Qin}, {Shen}, {Li}, {Mao}, {Smith}, {Rich},
  {Kunder}  \& {Liu}}{{Qin} et~al.}{2015}]{shen.2015}
{Qin} Y.,  {Shen} J.,  {Li} Z.-Y.,  {Mao} S.,  {Smith} M.~C.,  {Rich} R.~M.,
  {Kunder} A.,   {Liu} C.,  2015, \mn@doi [\apj] {10.1088/0004-637X/808/1/75},
  \href {https://ui.adsabs.harvard.edu/abs/2015ApJ...808...75Q} {808, 75}

\bibitem[\protect\citeauthoryear{{Richstone} \& {Tremaine}}{{Richstone} \&
  {Tremaine}}{1985}]{Richstone.1985}
{Richstone} D.~O.,  {Tremaine} S.,  1985, \mn@doi [\apj] {10.1086/163455},
  \href {https://ui.adsabs.harvard.edu/abs/1985ApJ...296..370R} {296, 370}

\bibitem[\protect\citeauthoryear{{Rodr{\'\i}guez} \&
  {Padilla}}{{Rodr{\'\i}guez} \& {Padilla}}{2013}]{Rodr.2013}
{Rodr{\'\i}guez} S.,  {Padilla} N.~D.,  2013, \mn@doi [\mnras]
  {10.1093/mnras/stt1168}, \href
  {https://ui.adsabs.harvard.edu/abs/2013MNRAS.434.2153R} {434, 2153}

\bibitem[\protect\citeauthoryear{{Rybicki}}{{Rybicki}}{1987}]{Rybicki}
{Rybicki} G.~B.,  1987, in {de Zeeuw} P.~T.,  ed.,  IAU Symposium Vol. 127,
  Structure and Dynamics of Elliptical Galaxies. p.~397,
  \mn@doi{10.1007/978-94-009-3971-4_41}

\bibitem[\protect\citeauthoryear{{Schwarzschild}}{{Schwarzschild}}{1982}]{Schwarzschild.19822}
{Schwarzschild} M.,  1982, \mn@doi [\apj] {10.1086/160531}, \href
  {https://ui.adsabs.harvard.edu/abs/1982ApJ...263..599S} {263, 599}

\bibitem[\protect\citeauthoryear{{Shen}}{{Shen}}{2014}]{Shen.2014}
{Shen} J.,  2014, in {Feltzing} S.,  {Zhao} G.,  {Walton} N.~A.,   {Whitelock}
  P.,  eds,  Vol. 298, Setting the scene for Gaia and LAMOST. pp 201--206,
  \mn@doi{10.1017/S1743921313006376}

\bibitem[\protect\citeauthoryear{{Shen}, {Rich}, {Kormendy}, {Howard}, {De
  Propris}  \& {Kunder}}{{Shen} et~al.}{2010}]{shen.2010}
{Shen} J.,  {Rich} R.~M.,  {Kormendy} J.,  {Howard} C.~D.,  {De Propris} R.,
  {Kunder} A.,  2010, \mn@doi [\apjl] {10.1088/2041-8205/720/1/L72}, \href
  {https://ui.adsabs.harvard.edu/abs/2010ApJ...720L..72S} {720, L72}

\bibitem[\protect\citeauthoryear{{Sheth}, {Vogel}, {Regan}, {Thornley}  \&
  {Teuben}}{{Sheth} et~al.}{2005}]{2005.Sheth}
{Sheth} K.,  {Vogel} S.~N.,  {Regan} M.~W.,  {Thornley} M.~D.,   {Teuben}
  P.~J.,  2005, \mn@doi [\apj] {10.1086/432409}, \href
  {https://ui.adsabs.harvard.edu/abs/2005ApJ...632..217S} {632, 217}

\bibitem[\protect\citeauthoryear{{Stark}}{{Stark}}{1977}]{Stark.1977}
{Stark} A.~A.,  1977, \mn@doi [\apj] {10.1086/155164}, \href
  {https://ui.adsabs.harvard.edu/abs/1977ApJ...213..368S} {213, 368}

\bibitem[\protect\citeauthoryear{{Valluri} \& {Merritt}}{{Valluri} \&
  {Merritt}}{1998}]{Valluri.1998}
{Valluri} M.,  {Merritt} D.,  1998, \mn@doi [\apj] {10.1086/306269}, \href
  {https://ui.adsabs.harvard.edu/abs/1998ApJ...506..686V} {506, 686}

\bibitem[\protect\citeauthoryear{{Valluri}, {Merritt}  \& {Emsellem}}{{Valluri}
  et~al.}{2004}]{2004.Valluri}
{Valluri} M.,  {Merritt} D.,   {Emsellem} E.,  2004, \mn@doi [\apj]
  {10.1086/380896}, \href
  {https://ui.adsabs.harvard.edu/abs/2004ApJ...602...66V} {602, 66}

\bibitem[\protect\citeauthoryear{{Valluri}, {Debattista}, {Quinn}  \&
  {Moore}}{{Valluri} et~al.}{2010}]{Valluri2.2010}
{Valluri} M.,  {Debattista} V.~P.,  {Quinn} T.,   {Moore} B.,  2010, \mn@doi
  [\mnras] {10.1111/j.1365-2966.2009.16192.x}, \href
  {https://ui.adsabs.harvard.edu/abs/2010MNRAS.403..525V} {403, 525}

\bibitem[\protect\citeauthoryear{{Valluri}, {Shen}, {Abbott}  \&
  {Debattista}}{{Valluri} et~al.}{2016}]{Valluri.2016}
{Valluri} M.,  {Shen} J.,  {Abbott} C.,   {Debattista} V.~P.,  2016, \mn@doi
  [\apj] {10.3847/0004-637X/818/2/141}, \href
  {https://ui.adsabs.harvard.edu/abs/2016ApJ...818..141V} {818, 141}

\bibitem[\protect\citeauthoryear{{Van den Bosch}, {van de Ven}, {Verolme},
  {Cappellari}  \& {de Zeeuw}}{{Van den Bosch} et~al.}{2008}]{Bosch.2008}
{Van den Bosch} R.~C.~E.,  {van de Ven} G.,  {Verolme} E.~K.,  {Cappellari} M.,
    {de Zeeuw} P.~T.,  2008, \mn@doi [\mnras]
  {10.1111/j.1365-2966.2008.12874.x}, \href
  {https://ui.adsabs.harvard.edu/abs/2008MNRAS.385..647V} {385, 647}

\bibitem[\protect\citeauthoryear{{Vasiliev}}{{Vasiliev}}{2013}]{Vasiliev.2013}
{Vasiliev} E.,  2013, \mn@doi [\mnras] {10.1093/mnras/stt1235}, \href
  {https://ui.adsabs.harvard.edu/abs/2013MNRAS.434.3174V} {434, 3174}

\bibitem[\protect\citeauthoryear{{Vasiliev}}{{Vasiliev}}{2019}]{Vasiliev.2019a}
{Vasiliev} E.,  2019, \mn@doi [\mnras] {10.1093/mnras/sty2672}, \href
  {https://ui.adsabs.harvard.edu/abs/2019MNRAS.482.1525V} {482, 1525}

\bibitem[\protect\citeauthoryear{{Vasiliev} \& {Valluri}}{{Vasiliev} \&
  {Valluri}}{2020}]{Vasiliev.2019c}
{Vasiliev} E.,  {Valluri} M.,  2020, \mn@doi [\apj] {10.3847/1538-4357/ab5fe0},
  \href {https://ui.adsabs.harvard.edu/abs/2020ApJ...889...39V} {889, 39}

\bibitem[\protect\citeauthoryear{{Virtanen} et~al.,}{{Virtanen}
  et~al.}{2020}]{Virtanen.2020}
{Virtanen} P.,  et~al., 2020, \mn@doi [Nature Methods]
  {10.1038/s41592-019-0686-2}, \href
  {https://ui.adsabs.harvard.edu/abs/2020NatMe..17..261V} {17, 261}

\bibitem[\protect\citeauthoryear{Wang, Mao, Long  \& Shen}{Wang
  et~al.}{2013}]{Mao.2012}
Wang Y.,  Mao S.,  Long R.~J.,   Shen J.,  2013, \mn@doi [Monthly Notices of
  the Royal Astronomical Society] {10.1093/mnras/stt1537}, 435, 3437

\bibitem[\protect\citeauthoryear{{Weinberg}}{{Weinberg}}{1985}]{1985.Weinberg}
{Weinberg} M.~D.,  1985, \mn@doi [\mnras] {10.1093/mnras/213.3.451}, \href
  {https://ui.adsabs.harvard.edu/abs/1985MNRAS.213..451W} {213, 451}

\bibitem[\protect\citeauthoryear{{Weiner}, {Sellwood}  \& {Williams}}{{Weiner}
  et~al.}{2001}]{Weiner.2001}
{Weiner} B.~J.,  {Sellwood} J.~A.,   {Williams} T.~B.,  2001, \mn@doi [\apj]
  {10.1086/318289}, \href
  {https://ui.adsabs.harvard.edu/abs/2001ApJ...546..931W} {546, 931}

\bibitem[\protect\citeauthoryear{{Zhao}}{{Zhao}}{1996}]{1996.Zhao}
{Zhao} H.,  1996, \mn@doi [\mnras] {10.1093/mnras/283.1.149}, \href
  {https://ui.adsabs.harvard.edu/abs/1996MNRAS.283..149Z} {283, 149}

\bibitem[\protect\citeauthoryear{{Zhu} et~al.,}{{Zhu} et~al.}{2018a}]{Zhu.2018}
{Zhu} L.,  et~al., 2018a, \mn@doi [Nature Astronomy]
  {10.1038/s41550-017-0348-1}, \href
  {https://ui.adsabs.harvard.edu/abs/2018NatAs...2..233Z} {2, 233}

\bibitem[\protect\citeauthoryear{{Zhu} et~al.,}{{Zhu}
  et~al.}{2018b}]{ling.1018}
{Zhu} L.,  et~al., 2018b, \mn@doi [\mnras] {10.1093/mnras/stx2409}, 473, 3000

\bibitem[\protect\citeauthoryear{{Zou}, {Shen}  \& {Li}}{{Zou}
  et~al.}{2014}]{Zou.2014}
{Zou} Y.,  {Shen} J.,   {Li} Z.-Y.,  2014, \mn@doi [\apj]
  {10.1088/0004-637X/791/1/11}, \href
  {https://ui.adsabs.harvard.edu/abs/2014ApJ...791...11Z} {791, 11}

\bibitem[\protect\citeauthoryear{{de Zeeuw} \& {Franx}}{{de Zeeuw} \&
  {Franx}}{1989}]{Zeeuw.1989}
{de Zeeuw} T.,  {Franx} M.,  1989, \mn@doi [\apj] {10.1086/167735}, \href
  {https://ui.adsabs.harvard.edu/abs/1989ApJ...343..617D} {343, 617}

\makeatother
\end{thebibliography}


\newpage	
	
\appendix


\section{Deprojection of more cases with different orientations } 
\label{ap1}
\begin{figure} 
	\centering
	\subfigure[][]{%
		\label{60-0}%
		\includegraphics[width=3.8cm]{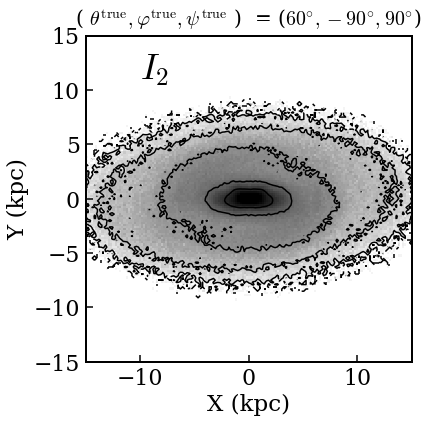}
		\includegraphics[width=4.8cm]{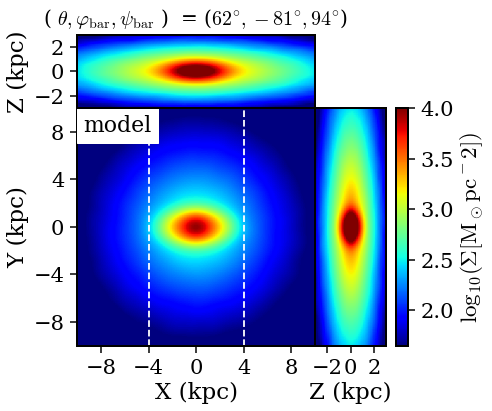}}\\
	\subfigure[][]{%
		\label{60-90}%
		\includegraphics[width=3.8cm]{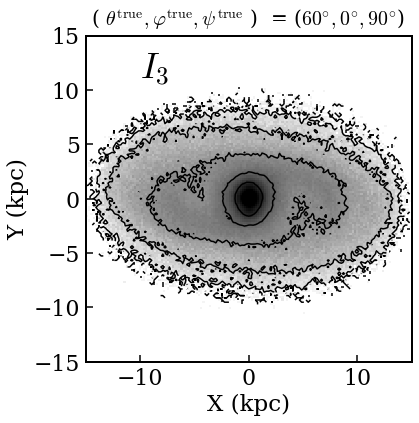}
		\includegraphics[width=4.8cm]{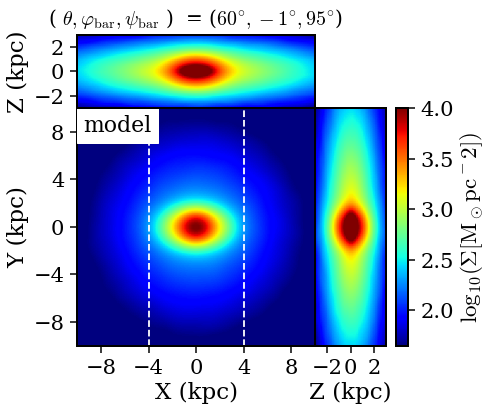}}\\
	
	\caption[]{ Deprojection of more mock galaxies, with different angles of $ \varphi_{\mathrm{bar}} $:
		\subref{60-0}) Mock data $ I_{2} $;
		\subref{60-90}) mock data $ I_{3} $. Dashed lines mark the full length of the bar in  simulation ($ \sim 8 \hspace{0.1cm} \mathrm{kpc} $). }%
	\label{all-mock}%
\end{figure}

\begin{figure} 
	\centering
	\subfigure[][]{%
		\label{80-0}%
		\includegraphics[width=3.8cm]{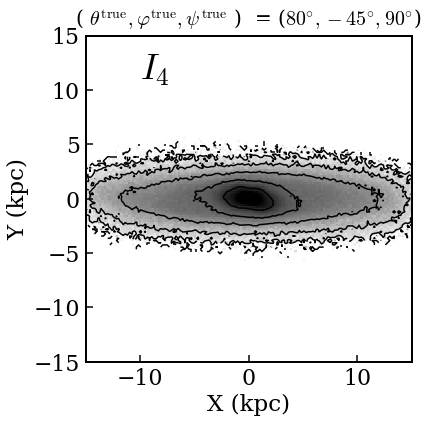}
		\includegraphics[width=4.8cm]{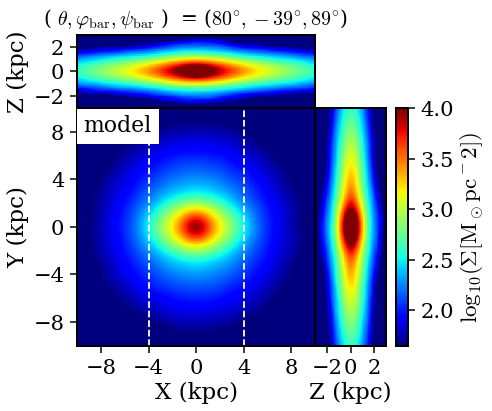}}\\
	
	\subfigure[][]{%
		\label{80-30}%
		\includegraphics[width=3.8cm]{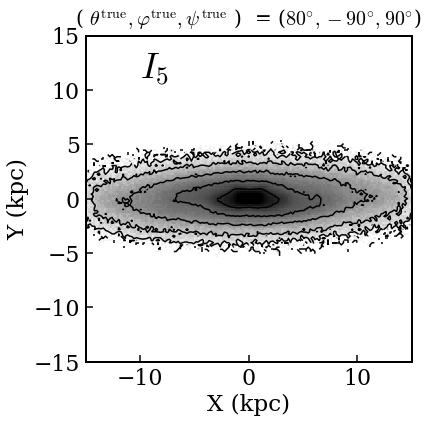}
		\includegraphics[width=4.8cm]{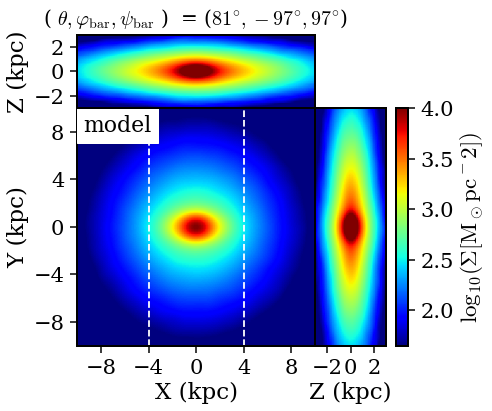}}\\%
	
	\subfigure[][]{%
		\label{80-90}%
		\includegraphics[width=3.8cm]{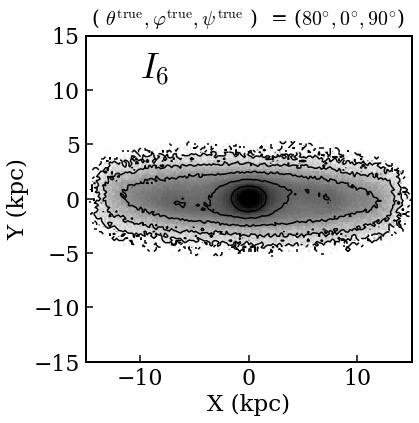}
		\includegraphics[width=4.8cm]{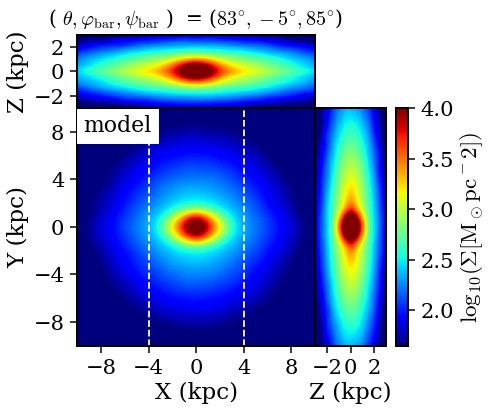}}%
	
	\caption[]{ Deprojection of edge-on mock galaxies with different angles of $ \varphi_{\mathrm{bar}} $:
		\subref{80-0}) Mock data $ I_{4} $ ;
		\subref{80-30}) mock data $ I_{5} $ ;
		\subref{80-90}) mock data $ I_{6} $.  Dashed lines mark the full length of the bar in simulation ($ \sim 8 \hspace{0.1cm} \mathrm{kpc} $).}%
	\label{all-mock2}%
\end{figure}

To show the effectiveness of our method for galaxies with various observational orientations, we obtain the 3D density maps for mock galaxies with different viewing angles as listed in Table \ref{table:mockt}. The results are shown in Figs. \ref{all-mock} and \ref{all-mock2}. The first and second columns show the mock images and deprojected density maps respectively. The 3D density inferred for these galaxies generally match the true 3D density distribution of the simulation well, in a similar manner to $ I_{1} $ which we described in detail.   

\section{Unmatched orbits} 
\label{ap2}
\begin{figure*}%
\centering
	\subfigure[][]{ %
		\label{fig:ap2A}%
\includegraphics[width=7.cm]{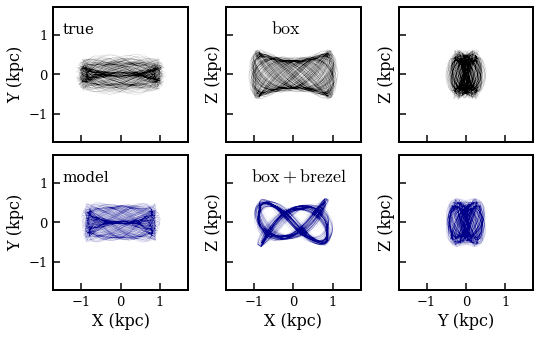}}
\subfigure[][]{ 
		\label{fig:ap2B} 
\includegraphics[width=7.1cm]{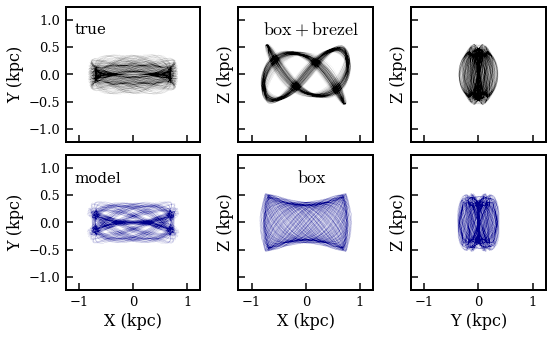}}
\subfigure[][]{ 
		\label{fig:ap2C} 
\includegraphics[width=7cm]{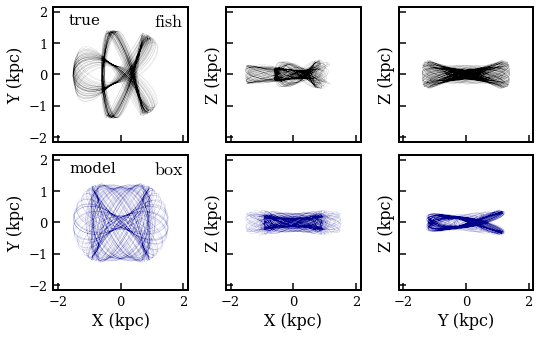}}%
\subfigure[][]{ 
		\label{fig:ap2D} 
\includegraphics[width=7cm]{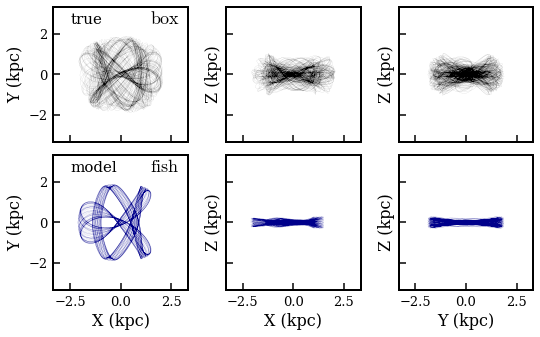}}
\subfigure[][]{ 
	\label{fig:ap2E} 
	\includegraphics[width=7cm]{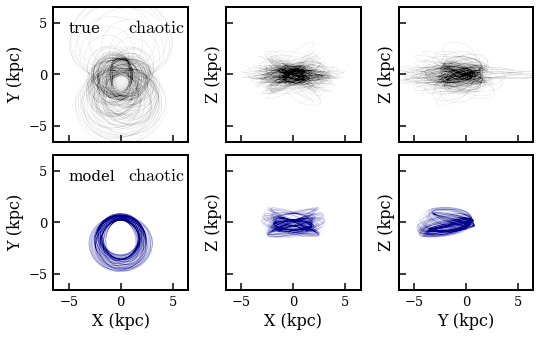}}%
\subfigure[][]{ 
	\label{fig:ap2F} 
	\includegraphics[width=7cm]{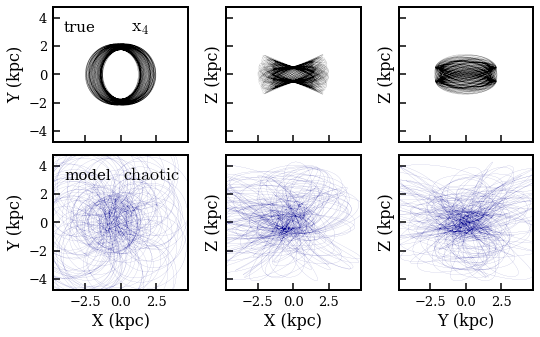}}
\
\caption[]{Typical unmatched orbits in the true potential (black), and in our model (blue), plotted in the $ x-y $, $ x-z $, and $ y-z $. Each pair of orbits is calculated with the same initial conditions. Resonant orbits and chaotic orbits are easy to be unmatched, they are sensitive to the minor changes of the potential. }%
	\label{fig:ap2}%
\end{figure*}

Fig. \ref{fig:ap2} shows some typical unmatched orbits. The first row shows the orbits which are boxes in $ x-y $ plane under the true and model potentials, while in $ x-z $ plane they have different shapes. In Fig. \ref{fig:ap2A} our model potential generates a brezel orbit while it happens inversely in Fig. \ref{fig:ap2B}. Also in Fig.  \ref{fig:ap2C} and \ref{fig:ap2D} the fish orbit is generated once in our model and once in the true potential for different starting points.
The third row shows the orbits which have different trajectories in all planes. Around $ 1/3 $ of visually chaotic orbits follow totally different trajectories in the true and model potentials, Fig. \ref{fig:ap2E} shows such an example. Overall most of the unmatched orbits are from chaotic or resonant orbits.

\section{Chaotic orbits criterion}
\label{ap3}
\begin{figure}
	\centering	
	\includegraphics[width=6cm]{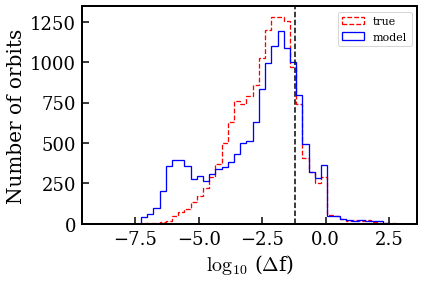}
	\caption[]{Distributions of frequency drift parameter $\log_{10} (\Delta f) $ for $ 15,000 $ orbits in the true (red) and model (blue) potentials. The dashed line indicates the threshold value of  $\log_{10} (\Delta f)= -1.2$. Orbits with $\log_{10} (\Delta f) >-1.2$ are classified as chaotic orbits.}%
	\label{df}%
\end{figure}

We examined if $\log_{10} (\Delta f) >-1.2$ used in \cite{Valluri2.2010, Valluri.2016} is a fair criterion to define chaotic orbits in our potential. In Fig. \ref{df}, we show the histogram distribution of  $\log_{10} (\Delta f) $ for the orbits in our model and in the true potential, which are similar to each other, especially at the high end with $\log_{10} (\Delta f) >-1.2$. We thus conclude it is a fair criterion. 

\section{measurement of the bulge-dominated region}
\label{ap4} 
\begin{figure}
	\centering	
	\includegraphics[width=6cm]{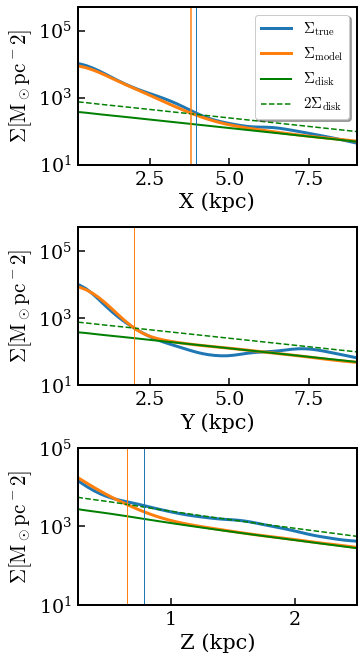}
	\caption[]{ Surface density distribution of the simulation (blue) and our model for mock galaxy $ I_{1}$ (orange); along the major axis (top), intermediate axis (middle), and minor axis (bottom). The green solid line and dashed line indicate $ \Sigma_\mathrm{disk} $ and $ 2 \times \Sigma_\mathrm{disk} $ of our model. We define the bulge-dominated region at the radius where $ \Sigma_\mathrm{total} $ = $ 2 \times \Sigma_\mathrm{disk} $. The blue and orange vertical lines mark the size of the bulge-dominated region in the simulation and in our model, respectively. }%
	\label{1Ddens}%
\end{figure}

  Fig. \ref{1Ddens} shows the surface density distribution along the three principal axes, blue is the simulation, and orange is our model inferred density from mock galaxy $ I_{1} $.
\par It is not straightforward to determine the bar size in three directions for a given density. To quantitatively compare the bulge-dominated region in simulation and in our model inferred density, we use the surface density of the disk in our model as a reference. We define the bulge-dominated region along each direction as the positions where $ 2 \times \Sigma_\mathrm{disk}=\Sigma_\mathrm{total} $. We obtain $(x_\mathrm{bulge},  y_\mathrm{bulge}, z_\mathrm{bulge})$ as $(\sim 4 \hspace{0.1cm} \mathrm{kpc}, \sim 2 \hspace{0.1cm} \mathrm{kpc}, \sim 0.78 \hspace{0.1cm} \mathrm{kpc})$ for the simulation and $( \sim 3.85 \hspace{0.1cm} \mathrm{kpc}, \sim 2 \hspace{0.1cm} \mathrm{kpc}, \sim 0.65 \hspace{0.1cm} \mathrm{kpc})$ for our model inferred density. 

\bsp	
\label{lastpage}

\end{document}